\documentstyle[11pt,aaspp4,tighten]{article}

\def\civ{C\,{\sc iv}}
\def\mg2{Mg\,{\sc ii}}
\def\mabs{$M_{abs}$}
\def\prad{$P_{rad}$}
\def\ks{$K_s$}
\def\rks{$r$$-$$K_s$}

\def\hk{$H$$-$$K$}
\def\jk{$J$$-$$K$}

\def\etal{{\it et\,al.}}
\def\ebv{$E$($B$$-$$V$)}

\def\q0{$q_0$}
\def\h0{$H_0$}
\def\hm1{$h^{-1}$}
\def\hn1{$h_{75}^{-1}$}
\def\n05{$N_{0.5}$}
\def\no5{$N_{0.5}$}

\def\z{$z$}
\def\h{$h$}

\def\Kuk{$K_{UKIRT}$}

\def\Kz{$K$$-$$z$}
\def\th{$\theta$}

\lefthead{HALL, GREEN AND COHEN}
\righthead{RLQ ENVIRONS I.}
\slugcomment{Accepted to The Astrophysical Journal Supplement May 26, 1998}

\begin{document}

\title{AN OPTICAL/NEAR-INFRARED STUDY OF RADIO-LOUD QUASAR ENVIRONMENTS I. METHODS AND $z$=1--2 OBSERVATIONS}

\author{Patrick B. Hall
\footnote{Visiting Student, Kitt Peak National Observatory, National Optical Astronomy Observatories, operated by AURA Inc., under contract with the National Science Foundation.}
\footnote{Current address:  Department of Astronomy, University of Toronto, 60 St. George Street, Toronto, Ontario, Canada M5S~3H8}}
\affil{Steward Observatory, The University of Arizona, Tucson, Arizona 85721 \\
Electronic Mail: hall@astro.utoronto.ca}
\author{Richard F. Green}
\affil{National Optical Astronomy Observatories, Tucson, Arizona 85726-6732\\
Electronic Mail: rgreen@noao.edu}
\author{Martin Cohen}
\affil{Radio Astronomy Laboratory, 601 Campbell Hall, University of California, 
Berkeley, California 94720\\ 
and Vanguard Research, Inc., Suite 204, 5321 Scotts Valley Drive, 
Scotts Valley, California 95066\\ Electronic Mail: mcohen@astro.berkeley.edu}

\begin{abstract}	\label{begin}

We have conducted an optical/near-infrared study of the 
environments of radio-loud quasars (RLQs) at redshifts \z=0.6--2.0.
In this paper we discuss the sample selection and observations for the
\z=1.0--2.0 subsample and the reduction and cataloguing techniques used.
We discuss technical issues at some length, since few detailed descriptions of 
near-IR data reduction and multicolor object cataloguing
are currently available in single literature references.

Our sample of 33 RLQs contains comparable numbers of flat-
and steep-radio-spectrum sources and sources of various radio morphologies,
and spans a similar range of $M_{abs}$ and $P_{rad}$, allowing us to 
disentangle dependence of environment on optical or radio luminosity from 
redshift evolution.

We use the standard ``shift-and-stare'' method of creating deep mosaiced
images where the exposure time (and thus the RMS noise) at each pixel
is not constant across the mosaic.  An unusual feature of our reduction
procedure is the creation of images with constant RMS noise from such mosaics.
We adopted this procedure to enable use of the FOCAS detection package over
almost the entire mosaic instead of only in the area of deepest observation 
where the RMS noise is constant, thereby roughly doubling our areal coverage.

We correct the object counts in our fields for stellar contamination using the
SKY model of Cohen (1995) and compare the galaxy counts to those in
random fields.  
Even after accounting for possible systematic magnitude
offsets, we find a significant excess of $K$$\gtrsim$19 galaxies.  Analysis and
discussion of this excess population is presented by Hall \& Green (1998).

\end{abstract}


\section{Introduction}	\label{intro}

The study of high-redshift galaxies and clusters is interesting because the 
light we see from them was emitted when galaxies and clusters were billions 
of years younger, and likely very different, than they are today.
Since deep field galaxy surveys have only begun to identify large numbers of
$z$$>$1 galaxies (\cite{ste96}), it is useful to seek other efficient methods
to find galaxies and clusters at \z$>$1.  

One such possible method is to look
for galaxies associated with quasars, specifically radio-loud quasars (RLQs).
Radio-quiet quasars (RQQs) are rarely found in clusters at any redshift, but
$\sim$35\% of intrinsically luminous (M$_{\rm B}$$<$$-$25) RLQs are located in
clusters of Abell richness class 0--1 (and occasionally 2) at $z$=0.5--0.7
(\cite{yg87}).  
However, little work has previously been done on RLQ environments at \z$>$0.7.

Some RLQs show possible additional evidence 
for being located in rich environments, in the form of an excess number of 
``associated'' \civ\ (\cite{fol88}) or \mg2\ (\cite{abe94}) absorption systems.
(For our purposes we define ``associated absorption'' to mean \civ\ or 
\mg2\ systems located within $\pm$5000~km~s$^{-1}$ of the quasar redshift.)
These systems may arise in gas expelled at high velocity from the quasars or 
in galaxies in clusters at or near the quasar redshifts.  
In the last few years 
high-resolution Keck spectra have shown that some associated 
\civ\ systems are unlike the typical intervening \civ\ absorption systems.
These ``intrinsic'' systems show unusually smooth and broad line
profiles, or well-resolved optically thick but shallow lines indicating partial
coverage of the background emission (\cite{hbj97}; \cite{ham97}).
Variability has also been seen in a few cases 
(\cite{ham95}; \cite{hbj97}; \cite{abf97}).
Associated \civ\ absorbers with these characteristics
are almost certainly produced by gas associated with the quasar central engine.
It is also possible that the excess associated \mg2\ absorbers of 
Aldcroft, Bechtold \& Elvis (1994) are intrinsic to their low-luminosity
steep-radio-spectrum sample, since it is the only sample to date to show 
an excess of associated \mg2.  
However, as the number of quasars with associated absorption which have been
studied in detail is small, it remains possible that a substantial fraction of
such quasars reside in clusters which produce associated absorption.

In 1994 we embarked upon a project to extend studies of quasar environments
beyond \z$\sim$0.6.  The goals of this project are to study the environments
of RLQs from $z$=0.6--2.0, to study the correlation, if any, between RLQ
environment and other quasar properties such as the presence of associated
absorption, and to study any examples of high-redshift galaxies and/or
clusters found in high-redshift RLQ fields.
To study galaxies at redshifts $z$$>$0.6 requires deep imaging, preferably in
the rest-frame optical and near-IR where galaxies emit most of their stellar
light.  Thus this project used \ks\ band (2.0--2.3$\mu$m) imaging to sample
the rest-frame near-IR at the quasar redshifts, supplemented by Gunn $r$ 
(0.6--0.7$\mu$m) imaging to sample the rest-frame near-UV.
Early-type galaxies which formed at $z$$\gg$2 will have very red colors in
$r-$\ks\ at $z$$>$1, which helps to distinguish any clustering around the 
quasars from the field galaxy population.

This paper discusses our observations for a study of RLQ environments at
$z$=1--2.  A \z=0.6--1 sample will be presented in a future paper.
In \S\ref{sample} we outline the selection of targets, in \S\ref{datared} we
discuss the observations and the data reduction and analysis techniques used,
and in \S\ref{systematics} we compare the galaxy counts in our fields to those
in random fields.  
In Paper 2 (\cite{hg98}) we examine the properties 
of galaxies in \z=1--2 RLQ fields, present the evidence for an excess population
of faint galaxies which are plausibly associated with the quasars, and
discuss the properties of these candidate high-redshift galaxies and clusters.

\section{Sample Selection}	\label{sample}

Our quasar sample was designed to cover the redshift range $z$=0.6--2.0 fairly
evenly, to contain comparable numbers of flat- and steep-radio-spectrum sources
and sources of various radio morphologies,
and to span a similar range of $M_{abs}$ and $P_{rad}$,
allowing us to study how environment correlates with such properties at a
given redshift and thus disentangle such effects from redshift evolution.
We preferentially selected objects with known high-$z$ intervening absorption
properties, giving even more preference to objects with few such absorbers,
to reduce confusion about with which system any detected excess
galaxies are associated.

\begin{figure}
\epsscale{0.8}
\plotone{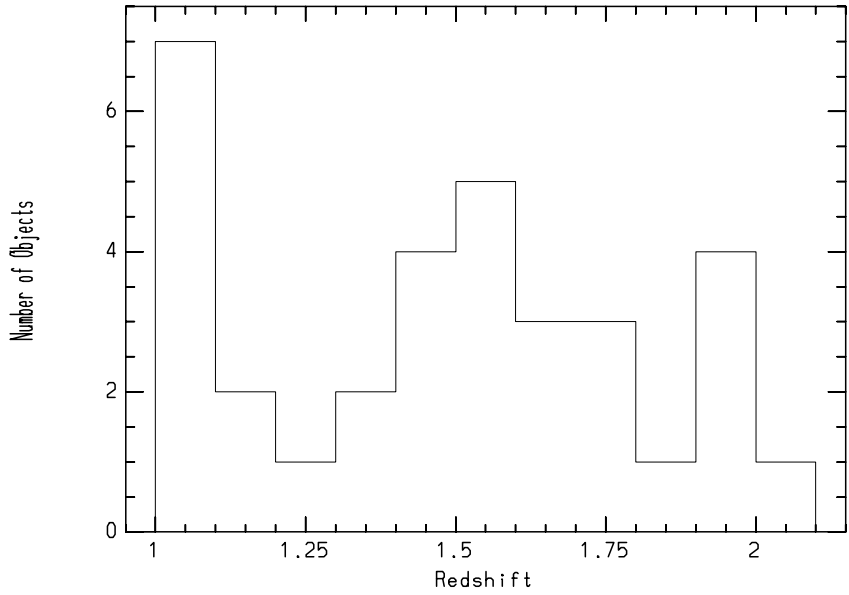}         
\caption[Redshift Histogram of $z$$>$1 Quasars]{
\singlespace
Redshift histogram of all $z$$>$1 quasars observed.
}\label{fig_zg1_z}
\end{figure}

The sample was split into three redshift ranges.
The $z$=0.6--1.0 subsample was intended to extend
previous quasar environment studies (which reached z$\sim$0.6--0.7) to the 
highest redshift feasible using the Steward 61$"$ and 90$"$ telescopes.
Data and analysis for this subsample will be presented in a future paper.
The $z$=1.4--2.0 subsample was selected to feasibly allow investigation of
intrinsic \civ\ absorption systems with the KPNO 4-m telescope.
The $z$=1.0--1.4 subsample was selected to link the 
low- and high-redshift samples to provide a view of quasar environments
over the entire redshift range $z$=0.6--2.0.
Figure \ref{fig_zg1_z} shows the redshift histogram of the $z$$>$1 quasars.

\subsection{Radio Properties}	\label{radio}

In order to study the dependence of environment on as many factors as possible,
in each redshift subsample we desired to span a similar range of $M_{abs}$ and
$P_{rad}$ and to have an even split between flat and steep 
radio spectrum sources and between objects of different radio morphologies.
Radio properties are given in 
Table \ref{tab_rad}.  

\begin{figure}
\epsscale{0.8}
\plotone{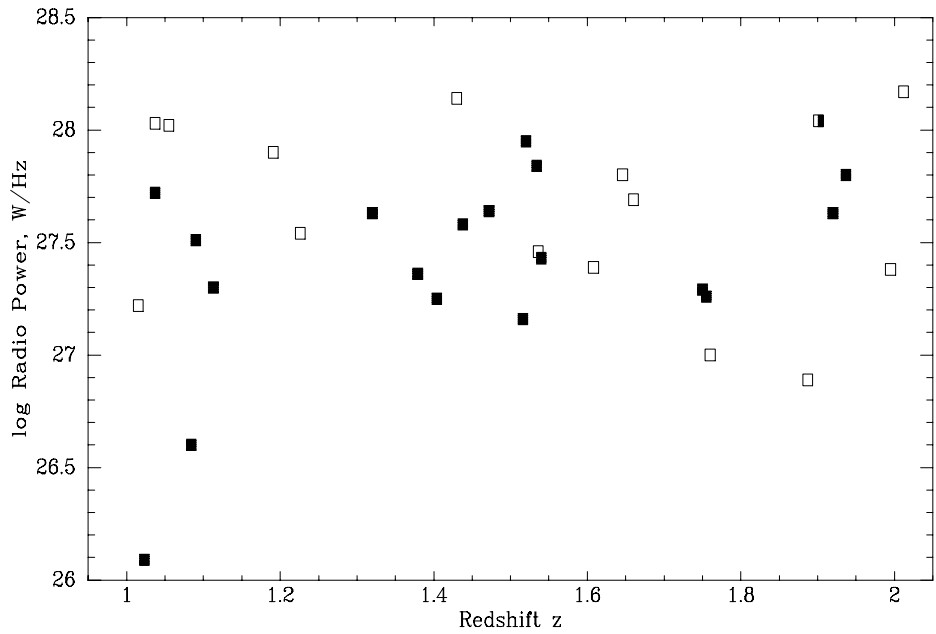}		
\caption[Radio Power vs. Redshift for $z$$>$1 Quasars]{
\singlespace
Radio Power log($P$) at 5~GHz rest frequency plotted vs. redshift for the 
$z$$>$1 quasars.  
Filled symbols are steep-spectrum objects, open symbols are flat-spectrum 
objects, and the half-filled symbol has unknown radio spectral slope.
}\label{fig_pradz}
\end{figure}

Steep radio spectrum sources were defined as having $\alpha_r$$\geq$0.5,
where $S_{\nu} \propto {\nu}^{-\alpha_r}$ (\cite{wb86}).
Radio spectral indices were taken from Stickel (personal communication)
or from the NASA/IPAC Extragalactic Database (NED).\footnote{The NASA/IPAC Extragalactic Database (NED) is operated by the Jet Propulsion Laboratory, California Institute of Technology, under contract to NASA.}
Radio flux densities $S_{\nu}$ at 5~GHz were also taken from NED, and were
converted to the radio power (luminosity) 
at 5~GHz rest frequency using 
$\alpha_r$ and assuming isotropic emission.
Figure \ref{fig_pradz} is a graph of \prad\ vs. $z$ for the $z$$>$1 objects.
The average \prad\ is 27.41$\pm$0.57 for the $z$=1--1.4 sample, and 
27.56$\pm$0.36 for the $z$=1.4--2.0 sample, so they are well matched.  However,
because radio emission is not always isotropic (e.g. in the case of beaming),
and again because radio variability affected both the measured $S_{\nu}$ and
$\alpha_r$, our values of \prad\ should be considered representative only.

\subsection{Moderate Redshift Subsample (1.0 $< z <$ 1.4)}  \label{sample_midz}

In this subsample we tried to exclude quasars with known associated and/or
intervening \mg2\ or \civ, although in practice not many objects in our sample
in this redshift range have been surveyed for absorption of either kind.  
Candidate objects for this redshift range were RLQs from the Large Bright
Quasar Survey (\cite{hoo95}), 
1~Jy RLQs (\cite{sk96}) with existing $R$ images obtained by Stickel,
and HST QSOALS Key Project targets (\cite{kir94}),
supplemented with RLQs from \cite{abe94}, \cite{ss92}, \cite{y91},
\cite{spi85}, \cite{dun89}, and \cite{jhb91} (see also \cite{jhb92})
which had little or no absorption along the line of sight.
Due to poor weather, the final sample of observed $z$=1.0--1.4 quasars
consists of only 12 objects:  5 flat-spectrum, 6 steep-spectrum, and 1 
of unknown radio spectral slope.  
Basic information on the objects is given in Table \ref{tab_zg1},
and radio properties in Table \ref{tab_rad}.
Information on intervening and associated absorption systems seen in the
quasars' spectra is given in Table \ref{tab_abs}.  
Three objects are known not to have associated absorption, and one is;
the rest have no published associated absorption information.

\subsection{High Redshift Subsample (1.4 $< z <$ 2.0)}	\label{sample_hiz}

This subsample is further divided in two: the ``absorbed'' subsample of
quasars with associated \civ\ absorption of rest-frame equivalent width (REW)
$>$1.5~\AA\ within 5000~km~s$^{-1}$ of the quasar emission redshift z$_e$
and the ``unabsorbed'' subsample of quasars without such absorption.
We selected targets from the unpublished Radio-Loud Survey of Foltz {\it et al},
targets with z$_a$$>$z$_e$ absorption as listed by \cite{jun88}, and targets
from \cite{fol86} and \cite{btt90}.  Targets were prioritized based on 
lack of high-redshift intervening absorption and having \mabs\ and
\prad\ in a range comparable to targets in the unabsorbed subsample.

The unabsorbed subsample contains quasars at $z$=1.4--2.0 with no 
\civ\ absorption of any strength within 5000~km~s$^{-1}$ of $z_e$.  Quasars
with known associated \mg2\ absorption were also excluded as a precaution.
Targets were selected from the catalogs of \cite{y91}, \cite{jhb91},
\cite{btt90}, and \cite{ss92}, based on the same criteria as the absorbed
subsample.  The presence of intervening \mg2\ absorption systems was considered
less problematic than \civ\ because the work of Steidel, Dickinson
\& Persson (1994) shows that intervening \mg2\ absorbers can be statistically
identified and excluded from the analysis, even at these high redshifts.  

The final sample of observed $z$=1.4--2.0 quasars is given in Table 
\ref{tab_zg1} 
and consists of 21 quasars:  8 flat-spectrum and 13 steep-spectrum;
14 with and 7 without associated \civ\ absorption.
Basic information on the objects is given in Table \ref{tab_zg1} 
and radio properties in Table \ref{tab_rad}.  
Table \ref{tab_zg1} also lists the coordinates of two control fields intended
for use with the $z$$>$1 quasar fields.  
Information on intervening and associated absorption seen in the quasars' 
spectra is given in Table \ref{tab_abs}.

\section{Observations, Data Reduction, and Object Cataloging}	\label{datared}

\begin{figure}
\plottwo{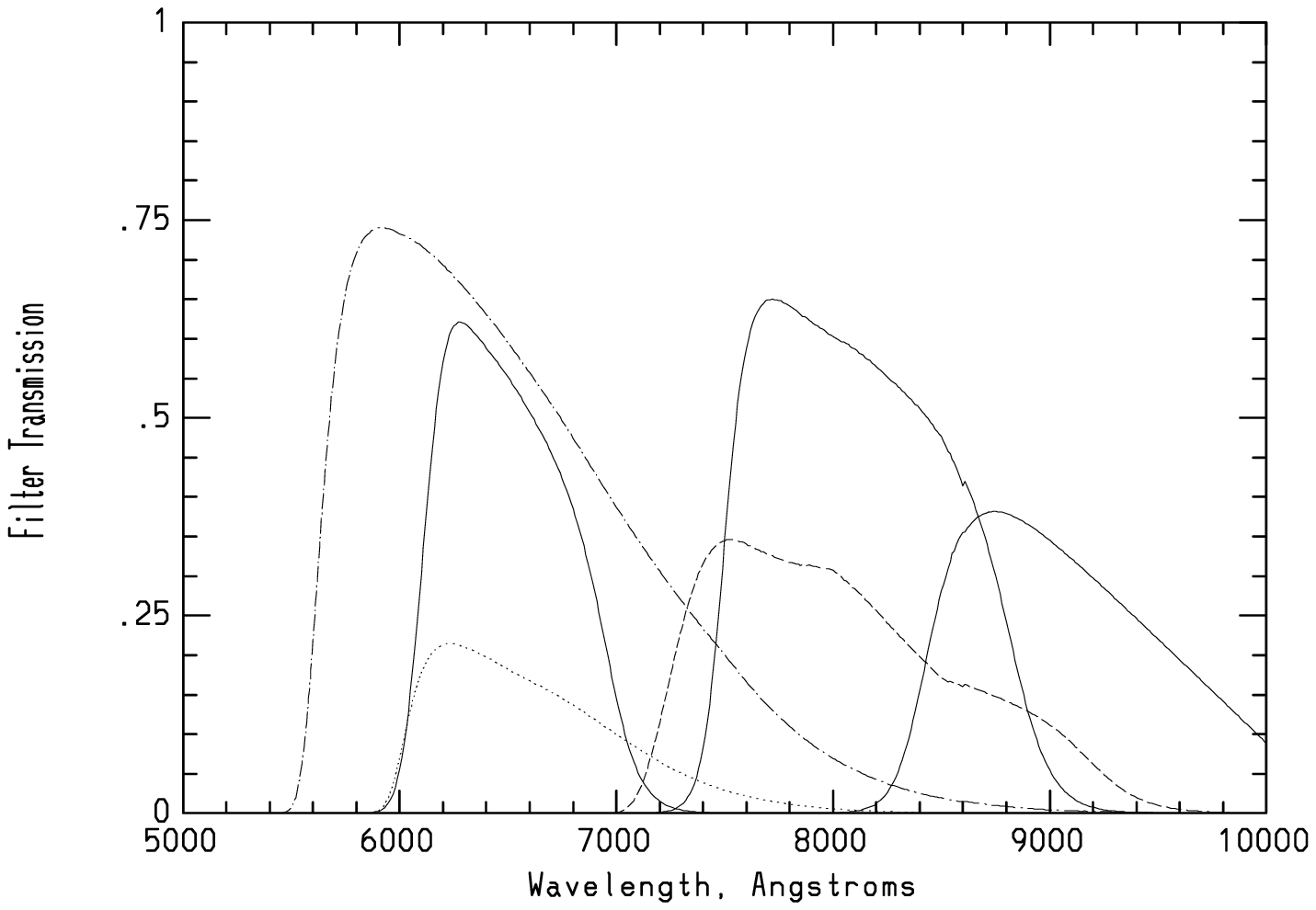}{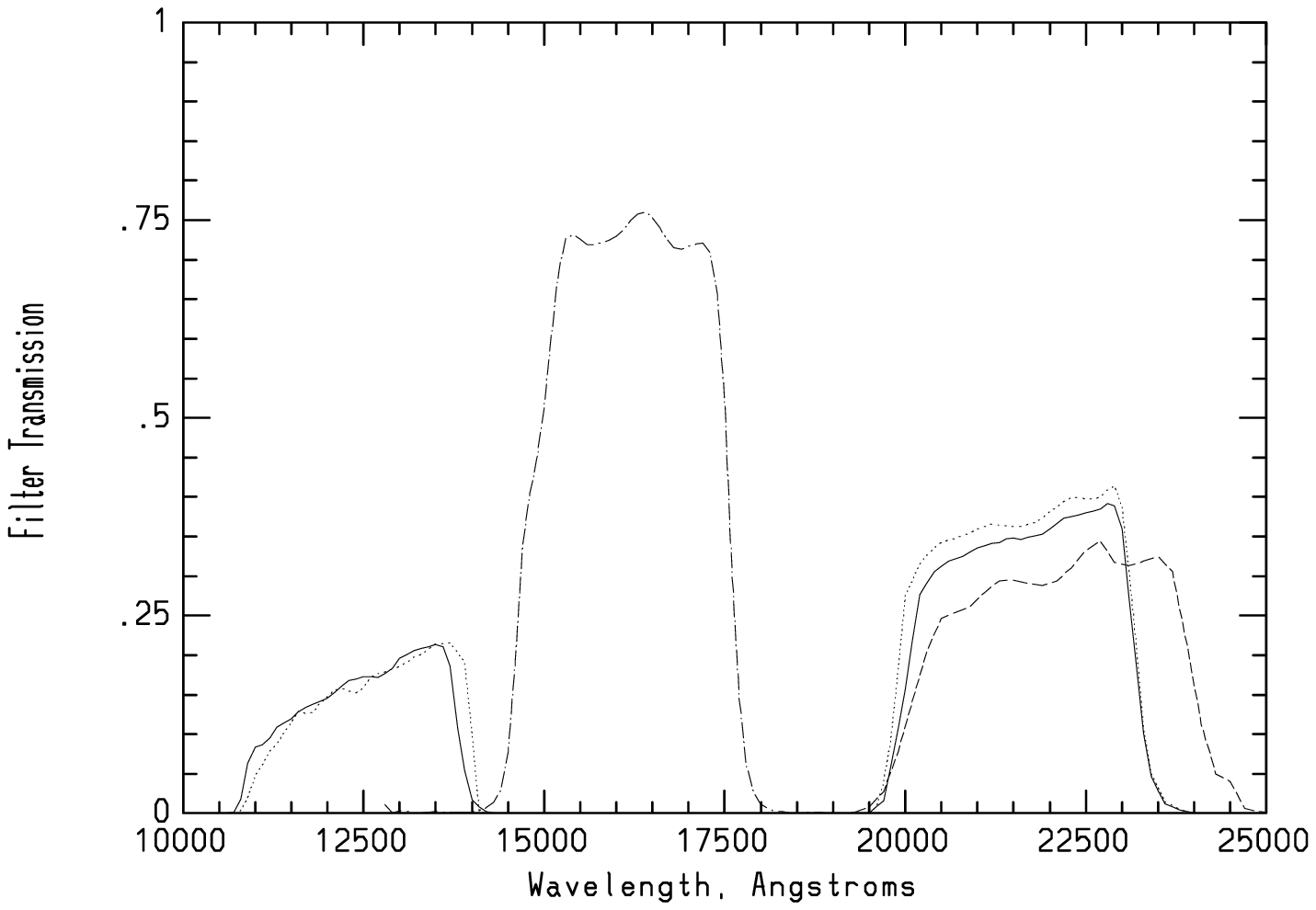}
\caption[Transmission of Filter Set and Instruments]{
\singlespace	
a. Transmission of all optical filters used, including detector QE.
Primary filters are Gunn $r$, $i$, and $z$ with Steward 800x1200 CCD (solid);
other filters used are Kron-Cousins $R_C$ with Steward 800x1200 CCD (dash-dot),
Kron-Cousins $I_C$ with Steward 2kx2k CCD (dashed),
and ``Osmer/Green'' $R'$ with old KPNO Tek2048 (dotted).  For clarity,
Gunn $riz$ with the lower-QE Steward 2kx2k CCD are not shown, although some
observations were made with those filter-instrument combinations.
b. Transmission of all IR filters used, including detector QE:  KPNO $J$ and 
IRTF $H$ with NSFCAM (dot-dash), and Steward $K$ with Steward 256x256 (dashed).
NSFCAM is an InSb array while IRIM and the Steward 256x256 are HgCdTe, which
accounts for the different throughput in $J$ and $K_s$ compared to $H$.
Note the wavelength scale is three times larger than in a.
}\label{fig_datared2}
\end{figure}

Optical observations were made in the Gunn {\it r, i, {\rm and} z} bands, 
with a handful of observations in Kron-Cousins $R_C$, $R'$ 
(see \S\ref{obsopt}), and 
Mould $I$, which is very similar to Kron-Cousins $I_C$ and is calibrated to it.
Near-infrared (used here to mean wavelengths from 1--2.5~$\mu$m) observations
were made in KPNO $J$, IRTF $H$, and 2MASS \ks\ (\cite{mcl95}) bands,
with a handful of observations in the Steward $J$ and $K$ bands.
Figure \ref{fig_datared2} shows the throughput of all filters
after accounting for the CCD or infrared array quantum efficiency.

\subsection{Near--Infrared Observations}	\label{obsir}

Observations of \z$>$1 targets and two moderately deep control fields
were made in the \ks\ and $J$ filters using the Kitt Peak National 
Observatory 4-meter Mayall telescope and Infra-Red Imager (IRIM),
a 256x256 NICMOS3 HgCdTe array
with gain of 10.46 e$^-$/ADU and read noise of 35~e$^-$ (\cite{pro95}).
The pixel scale was 0\farcs603 at \ks\ and 0\farcs608 at $J$.
There is a slight pincushion distortion in the IRIM field of view 
(Steidel, personal communication) for which no correction was made.

Images were typically taken in 4$\times$4 raster patterns with 15\arcsec\ steps,
$<$10\arcsec\ offsets between patterns, and 1 minute integrations per position.
Faint UKIRT standards (\cite{ch92}) were used for photometric calibration.
Minezaki \etal\ (1998a) found no color term between $K$ and \ks\ for the UKIRT
system, but see \S\ref{syslit}.

$H$-band images of Q~0835+580 and Q~1126+101 were
obtained through service observations on IRTF made by B. Golisch on UT 970319
in clear conditions.  NSFCAM was used with gain 10.0 e$^-$/ADU, read noise
55~e$^-$, and pixel scale 0\farcs3.
Two 4$\times$4 grids with 20$''$ steps and 6 coadds of 10 seconds exposure at
each position (32 minutes total) were made for each object.

\subsection{Near--Infrared Data Reduction}	\label{datared_ir}

To aid in the reduction of infrared array data, a collection of routines
collectively entitled {\sc phiirs} (Pat Hall's Infrared Imaging Reduction 
System) was developed to work within IRAF.\footnote{The Image Reduction and 
Analysis Facility (IRAF) is distributed by National Optical Astronomy 
Observatories, operated by the Association of Universities for Research in 
Astronomy, Inc., under contract to the National Science Foundation.}
{\sc phiirs} is available at the World Wide Web URL
{\it http://iraf.noao.edu/iraf/web/contrib.html}
or by contacting the first author.  
In general, ``standard'' data reduction techniques for infrared arrays were
used (e.g. Cowie {\it et~al.} 1990).
The specific routines used are in some cases directly derived from those in the
{\sc dimsum} package (Stanford, Eisenhardt \& Dickinson 1995) which is also 
available from the WWW URL given above.

\subsubsection{Outline}	\label{outline}

A brief outline of the steps used to reduce infrared array data follows.

1) Nonlinearity correction (IRIM data only).

2) Dark subtraction.

3) First-pass flattening, using a ``running skyflat" or domeflats.

4) First-pass sky subtraction, using a ``running sky".

5) Determination of image shifts and first-pass coadding.

6) Creation of individual-image object masks out of coadded image.

7) Second pass flattening (unless domeflats were used) and sky subtraction
using object masking and photometric scaling.

8) Cosmic ray removal (optional).

9) Make and apply throughput correction 
to account for light reaching the detector without being focussed by the
telescope (Steward 61$''$ and 90$''$ only).

10) Rotate and resample from different observing runs or telescopes to a common scale (if necessary).

11) Coadd second-pass (and possibly rotated) images with 2x2 pixel resampling.

12) Calculate appropriate scaling factors for nonphotometric data and re-coadd.

13) Remove low-level striping from coadded image, if necessary (IRIM only).

14) Photometric calibration.

\noindent{Where necessary, the reduction steps are now discussed in detail.}

\subsubsection{Nonlinearity Correction}	\label{nonlin}

To correct for nonlinearity in the KPNO IRIM array at high signal levels,
sequences of varying exposure time images of the illuminated flatfield screen 
was taken through a narrow 2.2$\mu$m filter during each run.
A fit was made to the observed dark-subtracted signal versus the
expected signal extrapolated from low light levels, accounting for the delay
between biasing and the first (non-destructive) read of the array, when the
array is accumulating charge which is not reflected in the final output value
for each pixel.
The fits were consistent between runs, so the results from all runs were
combined to find a nonlinearity correction which was applied to all images.

\subsubsection{Flattening and Sky Subtraction}	\label{flatsky}

The ``running sky" method of flattening and/or sky-subtracting was first
described by Cowie {\it et~al.} (1990).
Each image of the field, offset by a few arcseconds from the others,
is first flattened using a median of the raw images taken
immediately before and after it, and then sky-subtracted using a median 
of the flattened versions of the same images.  
(The order of these two operations can be reversed.)
Typically 8 images total, 4 before and 4 after, were used (20 or 22 for NSFCAM).

Running sky-subtraction was used for all data, but domeflattening was used for
IRIM and NSFCAM data.  
Using running skyflats alone or along with domeflats gave similar results of
RMS noises $\sim$10\% higher than domeflats alone.
This is probably because domeflats were typically constructed from a
set of images with total signal $\sim$20\% higher than the skyflats.

\subsubsection{Coadding Infrared Images} 	\label{coadd}

Extensive experimentation was done to determine the optimal parameters for 
coadding the data such that 
cosmic rays and other bad pixels were excluded,
the photometry of both bright and faint objects was not significantly 
different from a simple average,
and the RMS noise of the output image was as low as possible, to optimize the
detection of faint objects.
The weights used to make the final coadded image were the
``optimal weights" of each image: the exposure time divided by the variance 
(the square of the image's RMS noise, after iteratively rejecting cosmic
rays and other outlying pixel values).

The low S/N of individual IR data frames and occasional bad pixels makes the
interpolation required for fractional pixel shifting undesirable, but the large
number of offset images taken of each field does make subpixel information 
recoverable.
Thus each original pixel was typically replicated into four pixels (2$\times$2)
and the images coadded using integer shifts in units of these new pixels.

Our detection algorithm requires a constant RMS noise across the image
(see \S\ref{FOCAS}).  Thus along with each coadded image an exposure map was
created, giving the total exposure time at each pixel 
Assuming 
a constant sky background during the observations, at each pixel
the coadded image will have RMS noise proportional to (exposure time)$^{-1/2}$,
i.e. higher at the edges of the image where the exposure time was less.
By multiplying the square root of the exposure map image by the coadded image,
the coadded image is normalized to constant RMS noise.
However, when the sky background varies and the RMS noise does not scale 
as (exposure time)$^{-1/2}$, a more elaborate method must be used to construct
the normalization image; see \S\ref{optcoadd}.

\subsubsection{Object Masking}	\label{objmask}

When running flatfielding and/or sky-subtraction is used, small negative
residuals in the final coadded image are present around the position of each
such object, in the pattern of the dithering used to make the observations,
because faint objects are not completely excluded when the median is determined.
These residuals are eliminated by masking out objects detected in the
first-pass coadd and making a second pass.

To make this object mask image, each first-pass
coadded image was normalized to uniform pixel-to-pixel RMS using the
square root of the exposure map.  The normalized image was boxcar smoothed and
pixels above $\sim$5 times the smoothed-image RMS noise were flagged as objects,
along with rings of width one pixel around these flagged pixels.
Individual-image object masks were then copied from
the coadded-image mask using the known offsets of each image.
These masks were used in the second-pass flatfielding and sky-subtraction,
and the second-pass images were used to make the final coadd.

\subsubsection{Photometric Scaling} 	\label{photscale}

Since we wish to detect the faintest objects possible, we need to coadd many
infrared images, possibly taken at very different airmasses.
Thus it was decided to incorporate airmass corrections as multiplicative
scalings to the individual images before coadding.
Our \ks\ extinction coefficient was either 0\fm027 or 0\fm120 per airmass,
(see \S\ref{photkpno}),
so corrections applied to different images typically differ by $<$0\fm05
and always by $<$0\fm12.

As is usual, our magnitudes and fluxes represent exoatmospheric values,
i.e. values at zero airmass.
It should be noted that the extinction coefficient derived from observations at
airmasses $>$1 {\it underestimates} the extinction which occurs at airmasses
$<$1, i.e. in the upper atmosphere (\cite{joh65}; \cite{mb79}).
This is due to the Forbes effect:
light at wavelengths with large monochromatic extinction coefficients is
removed at small ($<$1) airmasses, leaving primarily light at wavelengths
with smaller monochromatic extinction coefficients at airmasses $>$1.
The Forbes effect is small in the optical 
but strong in the IR due to the many H$_2$O absorption lines.
To avoid potential systematic errors, carefully designed
photometric passbands should be used (\cite{yms94}).

Nonphotometric data were scaled to photometric data on the same field by
identifying several relatively bright objects (excluding the
quasars whenever possible) in the coadded image of all data on the field.
Photometry was then performed on these objects in the individual images,
discarding objects which fell on bad pixels.  
The relative scalings and weights for each image were then interactively 
examined and adjusted as necessary.  
Also, at this point, if necessary, data from different detector/telescope
combinations with different zeropoints were multiplicatively scaled to a 
common zeropoint.  This results in a higher weighting for data from more
efficient detector/telescope combinations.
The images were then re-coadded with these adjusted scalings and weights,
yielding a coadded image calibrated to the photometric data in the field.

\subsubsection{Destriping} 	\label{destripe}

On some 4-meter IRIM images, a pattern of ``striping" is evident.
This pattern is fixed on the sky but varies from field to field,
so it is thought to be caused by scattered light from bright stars well out of
the field of view (M. Dickinson, personal communication).  Subtraction of
smoothed versions of the object-masked images removed this pattern.
Running this procedure on images which showed no striping showed that it did
not introduce any systematic error in the photometry, but simply increased the
photometric uncertainties by at most $\sim$5\%.

\subsubsection{Rotation}	\label{rotate}

Instrument rotations relative to N-S on the sky
are not exactly the same between runs and telescopes.  Thus the images from
each separate filter or observing run were first coadded independently, and
these coadded images were used to determine the rotations between different
observing runs or filters, as well as the relative pixel scale between
different telescopes (see \S\ref{resample}).  Typically the rotation was
$\lesssim$1 degree, and the maximum observed value was $\sim$3 degrees.
One set of \ks\ images for each field was arbitrarily taken to have an
angle of zero relative to true north, so the angle of the coordinate systems
in these fields are expected to be good to only $\pm$1 degree RMS.

When images are rotated with linear interpolation, a gridlike pattern is
evident on the final image, caused by the differing noise characteristics
in regions where pixels are linear combinations of several original pixels.
Rather than rotating the final coadd and having such a pattern present,
cosmic-ray cleaned images were rotated with linear interpolation
before coadding, despite the low S/N of the individual frames, so that the
patterns on individual images were averaged over in producing the final coadd.

\subsubsection{Resampling}	\label{resample}

All data in $J$ and $H$ was resampled to the \ks\ pixel scale.
The final IR images have half the \ks\ pixel scale, or 0.3015\arcsec/pixel,
since they have been expanded 2x2 for better registration during coadding
(\S{\ref{coadd}}).
The resampling was done in the same step as the rotation, on the
cosmic-ray cleaned images.

\subsection{Optical Data Reduction}	\label{opt}

Observations in the Gunn {\it r} band were made for almost all z$>$1 targets
observed in the \ks\ band.
A few fields were observed in the $i$, $I$, and $z$ bands.
Standard reduction procedures for CCD data were used.  
Additional tasks for interactively removing fringes and for coadding optical
images in an accurate manner were developed to work within IRAF.  A 
package containing these tasks, entitled {\sc phat} (Pat Hall's Add-on Tasks),
is available at the WWW URL {\it http://iraf.noao.edu/iraf/web/contrib.html}
or by contacting the author.

The overall philosophy of the optical reductions is our desire for extracting
realistic magnitudes and errors over the maximum area and depth possible from
our imaging data.  
Extensive experimentation was done at all steps of the reduction process.
One important point is that we found no method to reliably coadd images 
with substantial seeing variations and reject cosmic rays without rejecting
valid pixels in the cores and/or wings of many objects and affecting photometry
at the $\sim$5\% level.  (McLeod (1994) also experienced this problem).
Thus we removed cosmic rays before coadding.

\subsubsection{Outline}	\label{optoutline}

A brief outline of the reduction steps for optical data follows.

1) Overscan subtraction.

2) Bias subtraction (and dark subtraction, if necessary).

3) Dome flattening.

4) Illumination correction using blank sky flats and/or twilight flats.

5) Removal of fringes and/or scattered light (if necessary).

6) Cosmic ray removal.

7) Reorient images from different observing runs, telescopes, or instruments
to a common orientation, and rescale to a common gain (if necessary).

8) First-pass coadding of images, using photometric scaling.

9) Rotate and rescale images from different observing runs or 
telescopes to match the coordinate system of the infrared images.

10) Calculate appropriate scaling factors for nonphotometric data and re-coadd.

11) Photometric calibration.

\noindent{Where necessary, the reduction steps are discussed in detail below.}

\subsubsection{Illumination Correction}			\label{illum}

Twilight flats were often used to improve the domeflattening.
For the Steward 800x1200 CCD, and occasionally the 2kx2k CCD, it was necessary
to also use sky flats from disregistered images taken at each position.
Even then, some 800x1200 images showed gradients of up to 5\%.  
The worst of these images were discarded, but in the interests of reaching
the faintest magnitudes, images flat to only a few percent were sometimes
used in producing the final coadds.  

\subsubsection{Fringe and Scattered Light Removal}		\label{fringescat}

Data taken in $z$ and some $i$ filters showed considerable fringing.  Fringe
images were created by medianing affected frames together and subtracting a
heavily smoothed background.  The fringe image was then scaled and subtracted
from the affected images, using an iterative procedure until satisfactory
results were obtained.
Bright scattered light was sometimes subtracted from individual object-masked
images using a similar procedure.  Other, fainter scattered light 
was masked out using the very useful IRAF task {\sc xray.ximages.plcreate}.

\subsubsection{Photometric Scaling}	 \label{optphotscale}

As with the infrared images, we chose to remove extinction by scaling the 
individual images to a common airmass rather than determining a mean airmass
for the final coadded image.  See \S\ref{photcal} for a discussion of the
photometric calibration and determination of the extinction coefficients.

\subsubsection{Coadding Optical Images}	 \label{optcoadd}

Unlike the near-IR images, where the FWHM of stellar images was 2--3 pixels
(at a pixel scale of 0.6\arcsec/pixel or larger),
in the optical images (0.3\arcsec/pixel) the FWHM was often 5 pixels or more.
The optical images also were taken over longer time spans, almost always
including data from different nights, resulting in
greater seeing variations between images.
No method could be found to reliably coadd images with substantial seeing 
variations and reject cosmic rays without rejecting valid pixels in the cores 
and/or wings of many objects and affecting photometry at the $\sim$5\% level
(McLeod (1994) also experienced this problem).
Such a method is possible when two or more images are taken at the same 
position, e.g. {\sc crrej} in {\sc stsdas}\footnote{{\sc stsdas} is distributed
by the Space Telescope Science Institute, which is operated by the Association
of Universities for Research in Astronomy, Inc., for the National Aeronautics 
and Space Agency.}.  
In principle this task could be adopted to cope with dithered images,
but for our optical images cosmic rays were always removed prior to coadding. 

The best method we found for removing cosmic rays was a slightly modified
version of the {\sc dimsum} task {\sc xzap} (\cite{sed95}).
The procedure was to subtract off a median-smoothed version of
the image, identify the peaks on the resultant image as cosmic rays, and 
replace them on the original image with the median of the surrounding pixels.
Typically a first pass was made using an object mask and a second pass without
object masking but with a more restrictive cosmic ray detection criterion.
Visual inspection and editing was done to remove the few obvious
cosmic rays which survived the automated removal.  The optical images were
then coadded with a simple averaging and minimal (or no) pixel rejection.

Another difference between coadding infrared and optical data arises from
the variable sky level in the optical.  The sky background is higher in the
infrared, but its variability at \ks\ is usually not extreme.  However, 
variable night sky emission lines are present in the $i$ and $z$ bands,
and the presence of the moon or of thin cirrus can affect the sky brightness in
the optical on a short timescale, a problem exacerbated by the fewer number
of optical images available for coadding.
Because the sky level (and thus the RMS) varies between individual images, 
the coadded-image RMS no longer scales with the exposure time at each pixel.
Thus instead of multiplying by the square root of the exposure time to 
normalize the image, it is necessary to multiply by the inverse of the expected
RMS for each pixel.  This normalization map was constructed by creating 
individual normalization images with a constant pixel value equal to the 
measured image variance, and then coadding together these images using a 
weighted average and offsets identical to those used to make the coadded image.

\subsection{Notes on Specific Optical Datasets}		\label{obsopt}

Optical observations of a few fields were obtained somewhat differently
from most fields.  
Fifteen minutes' $r$-band exposure on Q~1718+481 was obtained by B. Jannuzi
on UT 950702 using the Palomar 4m and COSMIC instrument.  
The zeropoint was found using an observation of HZ~44
and should be accurate to $\pm$5--10\%.
Three hours' exposure on Q~1258+404 was obtained in the Mould $I$-band
by J. Saucedo on UT 970322 and 970325 using the SO 90$''$ and 2kx2k CCD.
Photometric calibration was made to Kron-Cousins $I_C$ (actually Cape $I$;
see Sandage 1997) assuming a standard extinction coefficient of 0\fm061/airmass.
Without photometric observations of this field in Gunn $i$, we have not been
able to find a satisfactory transformation from $I_C$ to $i$, so we have left
the magnitudes as $I_C$.
Images of Q~2230+114 were obtained in Kron-Cousins $R_C$ by C. Liu 
on UT951222 and UT951225 using the SO 90$''$ and 800x1200 CCD.  
Conditions were nonphotometric, so this field cannot be used to study the
galaxy \rks\ color distribution.

Control field positions were selected from the Deep Multicolor Survey
(Hall {\it et~al.} 1996) 
with the sole requirement of having a bright spectroscopically confirmed star 
at their centers.  
Images were taken not in $r$ but in $R'$, a filter very similar to standard 
Kron-Cousins $R_C$ (and calibrated to it) but with less of a red tail.  
We converted $R_C$ magnitudes to $r$ using 
\begin{equation}
r = R_C + 0.322
\end{equation}
which was derived from both Frei \&
Gunn (1994) and Fukugita, Shimasaku \& Ichikawa (1995).

\subsection{Photometric Calibration}	\label{photcal}

Since observations at both optical and near-IR wavelengths were typically
made in only a single color, no color terms were used in the photometric
solutions for each filter.  The equation used to find the zeropoint and
extinction coefficients from standard star observations was the same
for all filters:
\begin{equation}		 \label{eq_photcal}
m = M - c_0 + c_1 \times X	
\end{equation}
where $m$ and $M$ are the observed (instrumental) 
and true (calibrated) standard star magnitudes respectively,
with $m$=$-$2.5$\times$log10($I$) where $I$ is the ADU/sec measured for the
standard,
$c_0$ is the zeropoint magnitude for 1~ADU/sec
(note that we use a different sign than is conventional),
$c_1$ is the extinction coefficient,
and $X$ is the airmass of the observation.

We express the zeropoint as a positive number so that each coadded image
$i$ with total exposure time $T_{exp}$ has a zeropoint $c_i$ written as:
\begin{equation}
c_i = c_0 + 2.5 \times {\rm log10}(T_{exp})
\end{equation}
The zeropoint $c_i$ is thus the magnitude of 1 ADU on the coadded image.
Thus the magnitude of an object on coadded image $i$ can be calculated as:
\begin{equation}
m_{obj} = c_i - 2.5 \times {\rm log10}(counts)
\end{equation}
where the object's counts are measured in ADU.  For fields with CCD images 
with different gains, images were multiplied by their gains before coadding.
The coadded image zeropoint was then adjusted by +2.5$\times$log10(gain) so
that the calibration was appropriate for the new gain=1 images.

\subsubsection{Steward 90$"$ + CCD}	\label{photccd}

Gunn $r$ (\cite{tg76}), $i$ (\cite{wad79}), and $z$ band (\cite{sgh83})
observations with the Steward 800x1200 and 2kx2k CCDs on the 90$"$ were
calibrated using standards from Thuan \& Gunn (1976), Wade {\it et~al.} (1979),
Kent (1985), J{\o}rgensen (1994), and Schneider (1995, personal communication).
For each observing run which was possibly photometric and had an adequate
number of standards, the photometric zeropoint and extinction coefficient
in each filter used were determined.  
When conditions were nonphotometric or sufficient data to determine the
extinction coefficients reliably were unavailable an $r$-band extinction
coefficient of 0.086 magnitude/airmass was assumed (\cite{ken85}).
The formal uncertainties on our 
magnitudes are 0\fm050 for $r$, 0\fm034 for $i$, and 0\fm038 for $z$.
The RMS scatters in each filter's photometric solution agree well with the
formal uncertainties.

\subsubsection{KPNO 4-meter + IRIM}	\label{photkpno}

For $J$ and \ks\ observations with IRIM on the KPNO 4-meter, photometric
calibration was performed using UKIRT faint IR standards (\cite{ch92}).
Data from all nights in each run were combined to solve for the photometric
zeropoint and extinction coefficient in each band separately
using the IRAF package {\sc photcal}.  

The February 1996 4-meter run had variable and often nonphotometric conditions.
It was the only 4-meter run where $J$ data was taken, so the
uncertainties in the $J$ calibration are slightly larger than for \ks\ or $H$.
The formal 1$\sigma$ uncertainty on our 
$J$ magnitudes is $\pm$0\fm063.
For \ks, the limited calibration data available for the February 1996 run were
consistent with the December 1994 run, and so the latter calibration was used
for both runs.

The \ks\ zeropoint was found to be 0\fm071$\pm$0\fm055 brighter during the 
July 1995 run than the December 1994 run,
possibly due to dust accumulation on the mirror.  
The extinction coefficients were also found to be different 
between the December 1994 and July 1995 runs.
Higher extinction in the summer is consistent with the findings
of Krisciunas {\it et~al.} (1987) for Mauna Kea and the predictions of
Manduca \& Bell (1979).
Both values of the extinction are plausible:  Stanford, Eisenhardt
\& Dickinson (1995) give 0\fm09/airmass for $K$ observations at KPNO,
equal to the average measured $K$ extinction coefficient for KPNO quoted in 
Manduca \& Bell (1979), whose calculations give 0\fm053 and 0\fm066/airmass 
for $K$ at KPNO during typical winter and summer conditions.
The formal uncertainties on our 
\ks\ magnitudes are $\pm$0\fm037 for 1994 and $\pm$0\fm054 for 1995,
but the RMS scatter between the two photometric solutions is $\pm$0\fm118.

No objects were observed in both runs to allow a direct check on the
\ks\ photometry, so must consider what systematic error might have been
introduced into our calibrated magnitudes if one or both photometric solutions
are in error.
In other words, our standard star observations indicated a difference in both
the telescope+instrument combination (zeropoint) and atmosphere (extinction
coefficient) between our 1994 and 1995 4m observing runs.  
However, in the extreme cases where the telescope+instrument, atmosphere,
or both were in fact exactly the same during the 1995 run as the 1994 run
and the difference in the photometric solutions is due to random error, 
we have introduced systematic offsets (and RMS uncertainties) of
\ks(1994$-$1995) of 0.071$\pm$0\fm055, 0.140$\pm$0\fm099, or 0.211$\pm$0\fm149,
respectively.  

We stress that it is the zeropoint $plus$ appropriate extinction uncertainties
which determine the uncertainties in the calibrated magnitude system,
rather than just the zeropoint uncertainties commonly quoted in the literature.
For a further discussion of possible systematic errors in our data,
see \S\ref{syswithin}.

\subsubsection{IRTF + NSFCAM}	\label{photirtf}

For the IRTF observations, three calibration observations of UKIRT standard
\#19 were made, all at airmass 1.2-1.22.
The zeropoint was found to be 22.034$\pm$0.057, 
in excellent agreement with the NSFCAM manual value of $-$22.06
(Leggett \& Deanult 1996).
The data were corrected 
using the $H$-band extinction of 0.051 magnitudes/airmass observed for Mauna Kea
by Krisciunas {\it et~al.} (1987).  
The formal uncertainty on our 
$H$ magnitudes is 0\fm060.

\subsection{Object Detection, Classification, and Photometry}	\label{FOCAS}

Prior to running object detection software on the final coadded images,
several important steps must be taken to ensure easy production and 
calibration of accurate output catalogs.

\subsubsection{Normalizing and Trimming the Coadded Images}	\label{trim}

First, the photometrically calibrated final coadded images of each field in each
filter are normalized to uniform RMS pixel-to-pixel noise using the exposure or
normalization maps, as appropriate (see \S\ref{coadd} and \S\ref{optcoadd}).
The images have already been rotated and resampled onto the same coordinate
system (see \S\ref{rotate}, \S\ref{resample} and \S\ref{optoutline});
they are now shifted to a common origin.
The images are then masked so that only areas with a certain minimum exposure
time (typically 0.25 times the maximum exposure time in the coadded image) are
included in the trimmed image.  Pixels outside such regions are set to zero.
For fields with images in multiple filters, the images are masked to exclude
pixels where any image has been set to zero.

For these trimmed images with uniform RMS, the detection 
significance of objects is constant across the image, i.e., a 3$\sigma$
detection is a 3$\sigma$ detection regardless of location on the image.
However, the magnitude scale and thus the limiting magnitude is not constant 
across the image:
the magnitudes of objects near the image edges will be erroneously faint,
since the edges have higher noise and have been
multiplied by some factor $<$1 to achieve constant RMS.
This complication is worth overcoming, as this procedure increases the useful 
area of our coadded images by about a factor of 2 compared to the standard 
procedure of using only the area with maximum exposure, because of the
small sizes of the IR arrays and CCDs used.
Glazebrook {\it et~al.} (1994) and Bershady, Lowenthal \& Koo employed similar
techniques for object detection, although not for photometry.
Another approach to the same problem is to modify object detection programs 
to handle spatially varying RMS noise, e.g. FOCAS (\cite{as96}) or SExtractor
(\cite{non98}).

Once a catalog of objects is generated, the true magnitudes are computed
by taking the central pixel of each object, determining the factor by
which that pixel was multiplied to make the normalized image,
and correcting the object's magnitude to account for this factor.
A similar procedure is used to calculate the area of sky surveyed as
a function of limiting magnitude.

\subsubsection{Summing Images in Different Filters}	\label{sum}

We are interested in galaxies detected in even just a single filter and wish to
study such galaxies' properties consistently (i.e. within the same aperture)
in all filters.
In particular, we are interested in objects in the $z$$>$1 quasar fields
detected only in the near-IR.
Thus we created a summed $r$+\ks\ image (or $r$+$J$+\ks\ where good $J$ data
was available) in each field.  
To give each filter's image equal weight at the faintest magnitudes,
all the input images were normalized to the same RMS before summing.

It is possible that an object detected just at the detection limit in one
filter could fall below the detection limit in the summed image, since the 
summing would effectively just be adding noise to such a galaxy.  
We used the FOCAS task {\sc clean} to replace the isophotal area of all
catalogued objects in the individual filter images with random sky values,
and then inspected the images visually.  Typically only one or two faint 
($<$5$\sigma$) blue galaxy candidates were overlooked in each field.
The candidates were often smaller than the minimum area or were classified
as ``noise,'' and the number of such candidates was consistent with the number
of noise spikes seen by displaying the negative side of the sky histogram.

We chose not to smooth the images to the same PSF before summing and 
determing isophotes for photometry, despite its attractiveness for
matching isophotes in different filters
and reducing noise to assist in faint object detection.
Tests of smoothing using IRAF {\sc immatch.psfmatch} and FOCAS showed
that there was a systematic shift in the magnitudes of objects after smoothing,
such that the objects were apparently fainter in the smoothed image.  
The shift was magnitude-dependent:  fainter objects had a larger magnitude
offset.  This is understandable since smoothing will reduce noise,
and at the faintest levels objects are difficult to distinguish from noise.
It might be possible to avoid this bias for e.g. $>$5$\sigma$ objects by
using a smaller convolution kernel than the 15$\times$15 pixel 
($\sim$4.5\arcsec$\times$4.5\arcsec) box we used, or by simply matching the 
FWHM of different images instead of the complete PSF shape as we attempted.

\subsubsection{Object Detection}	\label{detect}

Object detection, classification, and photometry was performed with FOCAS
(\cite{val82a}; \cite{val82b}).  
We used the built-in smoothing filter and required all detected pixels (in the
smoothed image) to be at least 2.5 times the iterative image RMS above sky.
We required initial detections to have a minimum area of 21 pixels 
(1.9 arcsec$^2$), although during the splitting phase a reduced criterion of 
10 pixels was used to better separate overlapping objects.
These values are appropriate for our typical seeing conditions.
For fields where the 2.5$\sigma_{sky}$ limit resulted in too many spurious 
detections, we retroactively excluded objects less than 3$\sigma_{sky}$ 
above sky.  
For a handful of fields with 
spurious pixel-to-pixel correlations from destriping of subpixel resampling
and shifting, or with poor seeing, FOCAS was rerun with a larger mininum area.
These detection criteria adjustments were carefully chosen to 
ensure that they eliminated no obviously real objects down to our 5$\sigma$
detection limits, and a negligible number to our 3$\sigma$ limits.

\subsubsection{Object Classification and Star/Galaxy Separation}   \label{classify}

To isolate the galaxies in our fields from noise and stars as best as possible,
we broke object classification down into several steps.
We believe our approach is 
robust down to \ks=16.5--18.5 depending on the field.
However, we can also correct statistically for the expected star counts
fainter than these limits, as detailed in \S\ref{SKY}.

First, automatic FOCAS classification was done using a different PSF template
for each field.
Obvious classification errors (e.g. residual cosmic rays not classified as
noise) were corrected interactively.
Objects classified ``noise" and ``long" were then removed from the catalog,
leaving a catalog with objects classified as either unresolved
(star or fuzzy star) or resolved (galaxy or diffuse).

However, star-galaxy separation is only accurate above a certain S/N.
To discriminate stars from galaxies down to this limit
we use the resolution parameter {\tt R}
of Bernstein {\it et~al.} (1994), defined as {\tt R}=$m_c$$-$$m_t$ where $m_c$
is the FOCAS ``core'' magnitude, the highest flux found in any contiguous 3x3
pixel area, and $m_t$ is the FOCAS total magnitude (see \S\ref{phot}).
Both values are measured in the summed image of each field
and thus they do not represent real magnitudes in any one filter.
We plot {\tt R} vs. $m_t$ from one field in Figure \ref{fig_datared4}, with
objects classified as stars by FOCAS shown as crosses and all other objects as
points.  Stars form a locus of small {\tt R} at bright magnitudes which runs
into the galaxy locus at some magnitude dependent on depth, seeing, and pixel
scale.  We classify as stars objects brighter than this magnitude which have
an {\tt R} value below an upper limit defined such that all objects in the
bright star sequence are still classified as stellar.  This region is marked
with the solid lines.  

Lastly, seven fields (noted in Table \ref{tab_zg1datainfo}) had useful WFPC2 
snapshots available from the $HST$ archive.
For these fields we changed the {\tt R} parameter classifications where
necessary to match classifications determined by eye from the $HST$ data.
Such changes should have negligible effects on star or galaxy counts.

\subsubsection{Object Photometry}	\label{phot}

The isophotes from the summed-image FOCAS catalog for each field were used to
create catalogs for each individual filter,
so that each object is measured within the same area in all filters.  
FOCAS total magnitudes (see next paragraph) are not expected to be affected
by the small differences between PSFs in the summed and individual images;
We have verified this by running FOCAS directly on the individual images.
Images not used to make the summed image typically had worse seeing than
those which were.  Nonetheless, we still used the summed-image FOCAS total 
magnitude isophotes for these images.
The same tests as above showed scatter consistent with photometric errors alone 
and no systematic errors in these cases as well, except for the $J$-band
observations of Q~2345+061 (2\farcs68 seeing).

For our magnitudes we used FOCAS total magnitudes, 
which have been shown to be unbiased estimates of the true magnitudes of 
unresolved and Gaussian-profile objects (\cite{kew89}; \cite{ber94}; 
\cite{bam94}; \cite{nw95}; but see \cite{eso97}).
FOCAS total magnitudes are derived by growing each object's original
detection isophote until it doubles in size and then measuring the
flux above the local sky in this aperture.

We make a minor correction to the FOCAS total magnitudes for split objects,
whose total magnitudes are derived by apportioning the original object's 
total counts among the split objects according to their isophotal areas.  
This does not take surface brightnesses into account.
A bright star and a low surface brightness galaxy with equal isophotal
areas will have erroneous magnitudes (too low and too high, respectively)
if they were originally detected as one object and later split by FOCAS.
Also, the detection isophote can be rather bright for objects very close
together which were barely split by FOCAS, leaving considerable flux 
outside it to be apportioned with the same problems as above.

\begin{figure}
\epsscale{1.0}
\plottwo{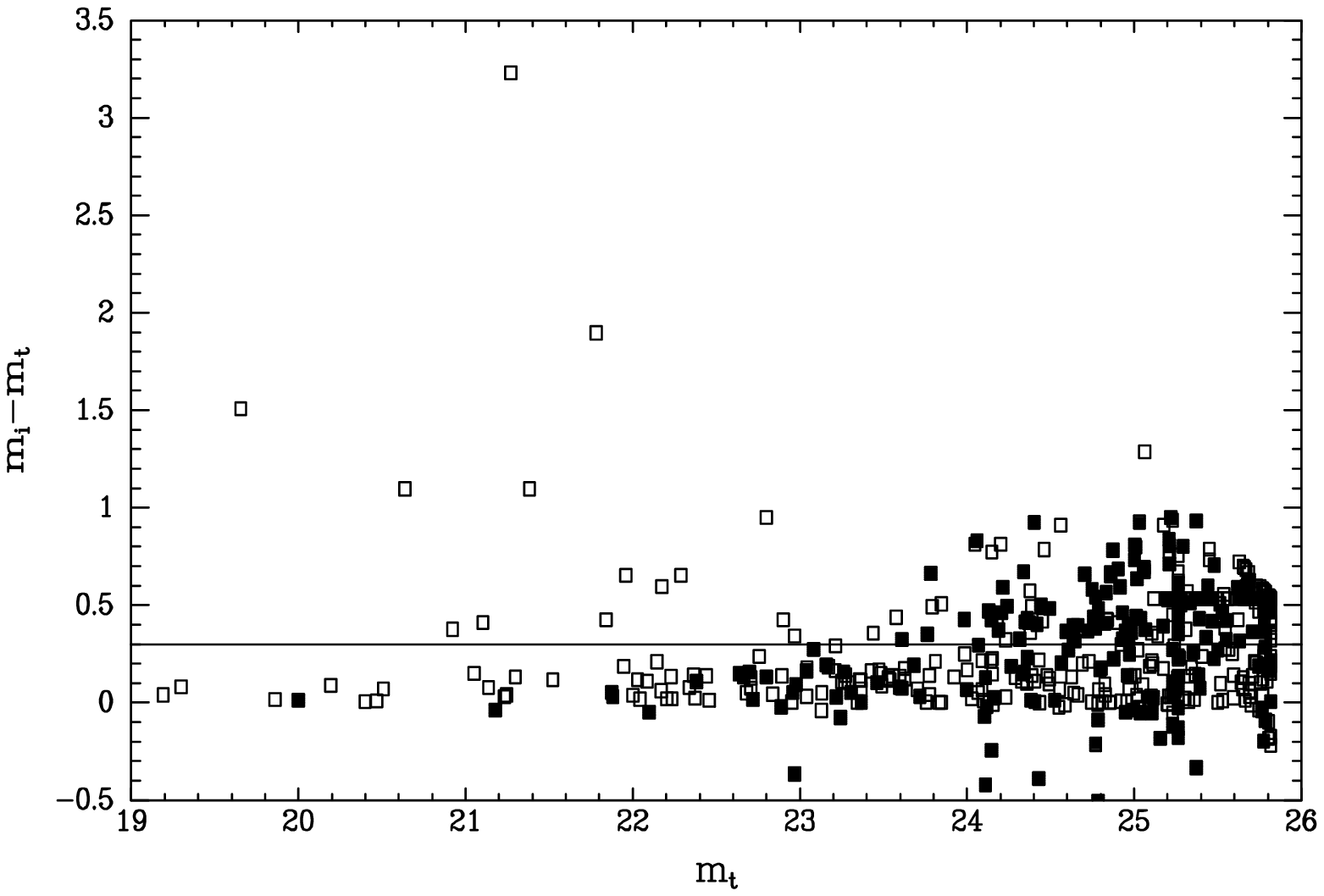}{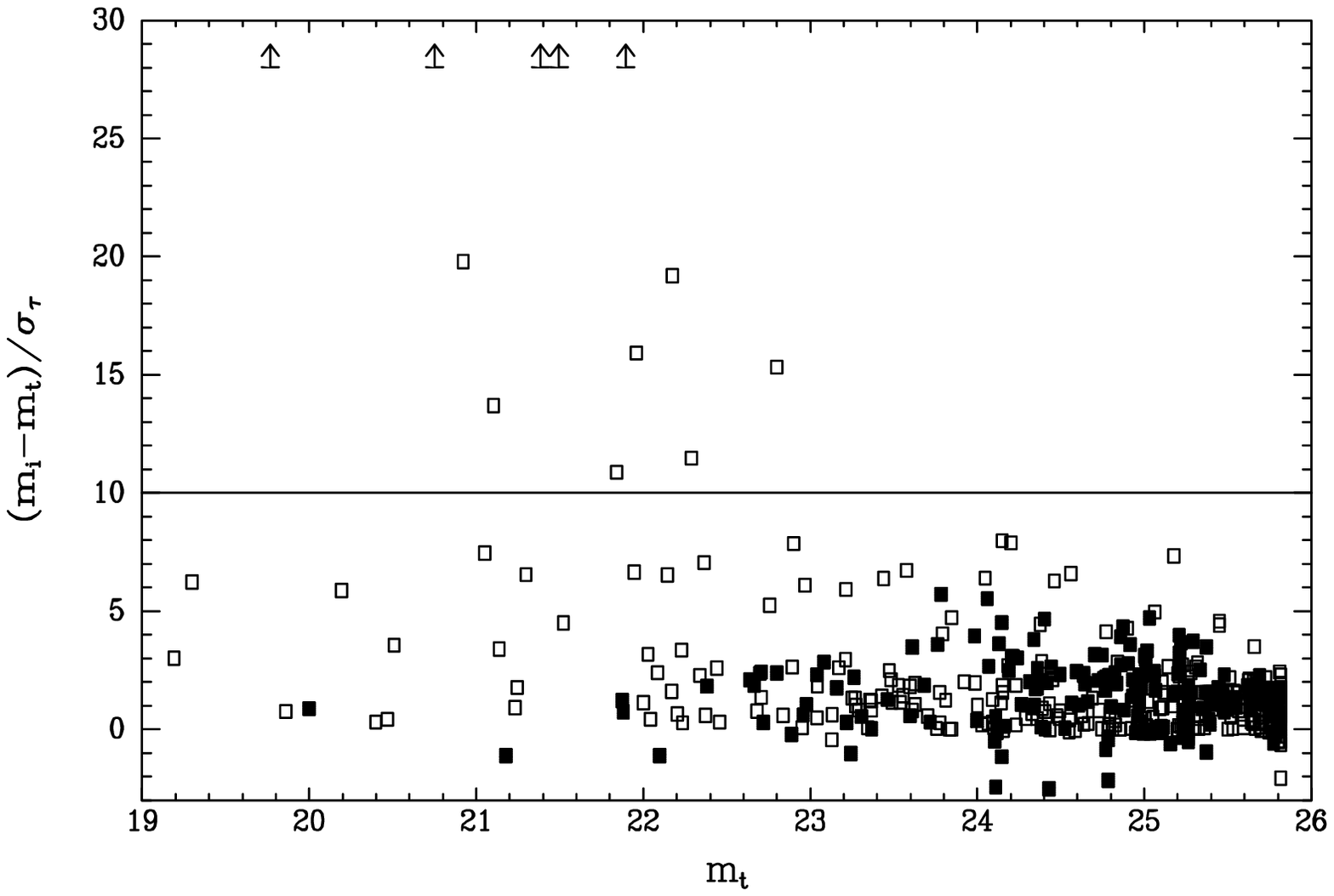}
\caption[Example of FOCAS Total Magnitude Correction Criterion]{
\singlespace
a) Aperture correction $m_i$$-$$m_t$ between total and isophotal FOCAS 
magnitudes plotted against FOCAS total
magnitude $m_t$ for all objects in the $r$ image of the Q~0835+580 field.
b) Uncertainty-normalized aperture correction ($m_i$$-$$m_t$)/$\sigma_t$ plotted
against $m_t$.  Lower limits indicate objects with values off the graph.
Objects never split by FOCAS are shown as filled boxes, and open boxes
are objects split by FOCAS.
Objects above the lines in both diagrams are considered to have erroneous
$m_t$, as are objects with $m_i$$-$$m_t$$<$0.
}\label{fig_datared5}
\end{figure}

\begin{figure}
\epsscale{0.75}
\plotone{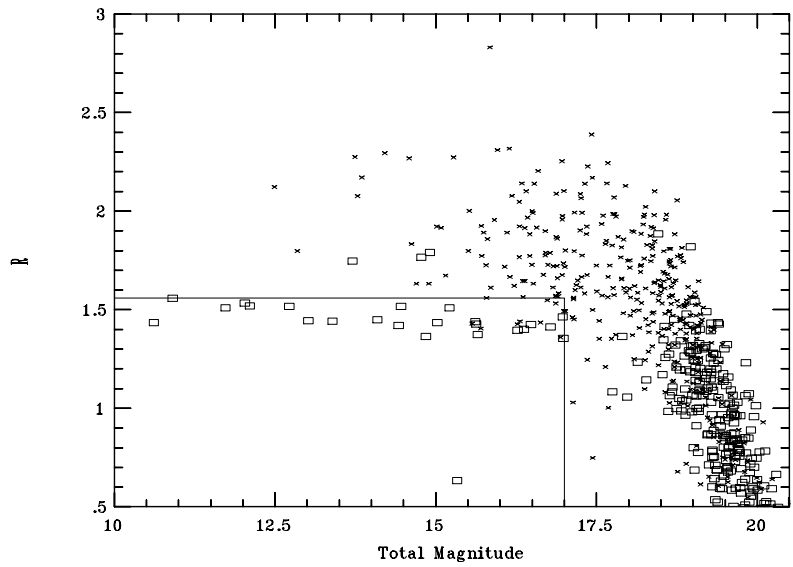}
\caption[Example of Star-Galaxy Separation Criterion]{
\singlespace
Resolution parameter {\tt R}=$m_c$$-$$m_t$ is plotted against FOCAS total
magnitude $m_t$ for all objects in the summed image of the Q~0835+580 field.
Objects classified as stars by FOCAS are shown as boxes and all other objects
as crosses.
Objects below and to the left of the lines are classified as stellar.
}\label{fig_datared4}
\end{figure}

To better estimate the total magnitudes of split objects, we examine the
difference between their total and isophotal magnitudes on the summed image,
$m_t$ and $m_i$.  As seen in Figure \ref{fig_datared5},
objects which were never split have low values of ($m_i$$-$$m_t$)/$\sigma_t$
where $\sigma_t$ is the uncertainty of the total magnitude $m_t$ 
(see next paragraph).
However, at bright magnitudes the uncertainties from photon statistics are
quite small, and so this quantity is sensitive to erroneous estimates of $m_t$.
Thus we empirically identified objects with 
($m_i$$-$$m_t$)/$\sigma_t$$>$10 and ($m_i$$-$$m_t$)$>$0\fm3 
as objects whose $m_t$ values are affected by splitting.
For these objects we replaced the total counts $c_t$ with
$c_t$=$c_i$+1.5$\times$$\Sigma_t$, where $c_i$ is the isophotal counts (in ADU)
and $\Sigma_t$ is the uncertainty of the FOCAS total magnitude, in ADU 
(see next paragraph).
As seen in Figure \ref{fig_datared5}, 
this approximates the average ``aperture correction'' of $\sim$1.5$\sigma_t$
between $c_t$ and $c_i$ for objects which were never split more accurately than
a fixed value 
would (compare Figure \ref{fig_datared5}a and b).

We have also replaced the total counts with the isophotal counts in cases where
the total counts were smaller, since the total magnitude should be brighter.
We thank the referee for pointing out that such cases are expected since the
total magnitude will have a larger uncertainty due to its larger aperture.
Thus our correction has introduced a statistical bias to brighter magnitudes.
However, only a small percentage of objects are affected
(see Fig.~\ref{fig_datared5}a) and so the estimated effect of the bias is an
erroneous brightwards shift in our magnitude bin centers of only 0\fm01--0\fm03.
This shift is not enough to significantly change any of our conclusions, so we
have not recalculated all our magnitudes at this time.  We plan to make our
catalogs public in a future paper, and will correct this error at that time.

After these corrections to the FOCAS total counts were made, 
total magnitudes were calculated from them.
FOCAS does not provide error estimates, so we calculated the error on the 
object counts $\Sigma_t$ and the magnitude error $\sigma_t$ as follows,
assuming a constant RMS noise across the image:
\begin{equation}
\Sigma_t = (g \times c_t + (g \times \sigma_{sky})^2 \times A_t)^{1/2}
\end{equation}
\begin{equation}
\sigma_t = 1.0857 \times \Sigma_t / (g \times c_t)
\end{equation}
where $g$ is the gain, $A_t$ the area of the FOCAS total magnitude aperture,
$c_t$ the total counts within that aperture, and $\sigma_{sky}$ the RMS noise
of the image (the original image, for images which showed striping).

The average 3$\sigma$ limiting magnitude was estimated for each filter from the 
magnitude-error graph before correction for RMS normalization (\S\ref{trim}).
Magnitudes fainter than this limit were set to the limiting value.
The true normalization-corrected magnitudes were then computed by determining
the factor by which the central pixel of the object was multiplied to make the
normalized image, and adjusting the object magnitude (or 3$\sigma$ limit)
appropriately.  
This assumes that the exposure time does not vary dramatically over the size
of a typical object.

\subsubsection{Galactic Extinction}	\label{galext}

We take the color excess \ebv\ from Burstein \& Heiles (1978, 1982)
and $R_{\lambda}$ from Mihalas \& Binney (1981).
We correct the limiting magnitudes and all objects' magnitudes for Galactic
extinction $A_{\lambda}$=$R_{\lambda}$$\times$\ebv\ in each filter.
Our adopted values of $R_{\lambda}$ are given in Appendix~\ref{photsys}
and the \ebv\ values for each field in Table~\ref{tab_zg1datainfo},
along with the 3$\sigma$ magnitude limits for each filter in each field.

\subsubsection{Construction of Overall Catalogs}	\label{catalog}

For each field, the individual filter catalogs were used to construct a catalog
of objects with a $\geq$3$\sigma$ detection within the FOCAS total magnitude 
aperture in either the $r$, \ks, or (when available) $J$ filters.
This is the faintest feasible catalog limit, since unresolved objects at the
3$\sigma$ limit have only a $\sim$50\% chance of detection (\cite{har90}).
We use our ``3$\sigma$ catalog" to compare N(m) counts with the literature.
However, for reliable comparative studies of galaxy colors and counts between
our different fields, this catalog was found to include too many spurious 
objects, since a spurious object in any filter becomes part of the catalog
whereas most real objects are common to all catalogs.
In addition, the errors on the colors become very large for objects below 
5$\sigma$.
Thus for most of our science we used a ``5$\sigma$ catalog," consisting of
all objects in the 3$\sigma$ catalog brighter than the average 5$\sigma$ $K_s$
limit in each field.  
This catalog is essentially magnitude limited; but as long as the 
exposure time at the edges is $\geq$36\% of the maximum, the average 5$\sigma$
$K_s$ magnitude limit will be greater than the 3$\sigma$ detection limit
across the entire field.

We now consider several points of importance in understanding the reliability
of our catalogs before making use of them scientifically.

{\em (1) Are our magnitudes and colors correct and on a system directly 
	comparable to published data?}
As mentioned in \S\ref{phot}, FOCAS total magnitudes have been shown to be
robust estimators of the true magnitudes of most types of faint galaxies, with
the exception of large, low surface brightness objects near the detection limit.
However, see \S\ref{systematics} for a discussion of possible 
systematic errors in photometric calibration in our work and the literature.

{\em (2) How many spurious objects are contained in the catalogs?}
We interactively removed obviously spurious objects in the halos and
diffraction spikes of bright stars.  
Faint cosmic rays may make some objects spuriously bright,
but the cosmic ray rate and our flux threshold for identifying them
are both low enough that this should be a neglible effect.  Misclassified stars
will contaminate the galaxy catalog below our star-galaxy classification limit.
The contamination from stars is relatively small at faint magnitudes,
but our fields are scattered over a wide range of Galactic latitude and
longitude (see Table \ref{tab_zg1}) 
and so stellar contamination cannot be ignored.  We address the statistical 
subtraction of faint star counts from our fields in \S\ref{SKY}.

We make no correction for truly spurious objects in our $5\sigma$ catalogs.
We reject objects classified as ``noise'' or ``long'' by FOCAS, and 
simulations (\cite{nwd91}) and observations by other researchers
(\cite{sma95}; \cite{hea97}) have shown that contamination by spurious
objects in single-filter data is $\lesssim$5\% at the 3$\sigma$ magnitude
limit for the default FOCAS detection threshold,
decreasing to essentially zero at the 5$\sigma$ limiting magnitude.
Thus we expect that essentially no $>$5$\sigma$ objects are spurious,
and that at most $\lesssim$10\% of objects in our 3$\sigma$ catalog with 
magnitudes fainter than the 5$\sigma$ limit are spurious ($\lesssim$15\% for
the four fields with $J$ data).

{\em (3) How many real galaxies do we miss, as a function of magnitude?}
Galaxies can be overlooked due to magnitude errors scattering them
below our detection limits, due to crowding, and due to misclassification as
noise or ``long'' objects by FOCAS or as bright stars by the {\tt R} parameter.
Misclassification is unlikely to exclude a significant number of galaxies
(see \S\ref{sum}).
Thus by the {\it completeness} of our galaxy catalogs we refer to
only the effects of crowding and magnitude errors.

The most sophisticated way to calculate the completeness is to calculate the
completeness matrix ${\bf C_{ij}}$ (\cite{ste90b}; \cite{mou97}; \cite{hea97}),
which gives the probabilities $P_{ij}$ that a galaxy of true magnitude $m_i$
is detected with magnitude $m_j$.
While this is still possible with our image normalization procedure, it is
rather complicated.  We opt instead to calculate just the completeness as a
function of magnitude $C(m)$, accounting for the variation in limiting 
magnitude across the image as follows.

At magnitudes well above our limiting magnitudes we assume that
the completeness $C(m)$ is affected only by crowding:
\begin{equation}
C(m) = {{A_{tot} - A(<m)}\over{A_{tot}}}
\end{equation}
where $A(<m)$ is the area on the sky covered by objects with magnitudes
brighter than $m$.  This conservatively assumes that a simulated object added
to the images would not be split by FOCAS if it overlapped an existing object.
At fainter magnitudes where it is possible that objects have been 
missed due to noise fluctuations,
we must account for the variation in limiting magnitude across the image.
We assume that completeness is a function of signal-to-noise only, ignoring its
dependence on surface brightness.
In each filter, we calculate the fraction $f$ of simulated objects recovered by
FOCAS as a function of counts in the normalized image, or equivalently as a
function of $m_{norm}$$=$z.p.$-$2.5$\times$log10(counts), the magnitude
measured on the normalized constant-RMS image.
(The technical details of this procedure are given in the following paragraph.)
We want to convert this to the recovery fraction $C(m_{real})$ as a function of
$m_{real}$, the real magnitude of an object after the normalization correction.
We know the area $A(\Delta m)$ of the image as a function of 
$\Delta m$=$m_{real}$$-$$m_{norm}$.
Thus we calculate $C(m_{real})$ by summing over the fractional area at each
$\Delta m$ multiplied by the fraction $f$ of recovered simulated objects with
a magnitude 
on the normalized image corresponding to that real magnitude:
\begin{equation}
C(m_{real})=\sum_{\Delta m} {{A(\Delta m)}\over{A_{tot}}} f(m_{real} - \Delta m)
\end{equation}
We also accounted for galactic extinction $A_{\lambda}$ in each filter of each
field so that the final completeness values
$C(m_{final})$=$C(m_{real}-A_{\lambda})$ were for identical values of
$m_{final}$ in each field.

\begin{figure}	
\plotfiddle{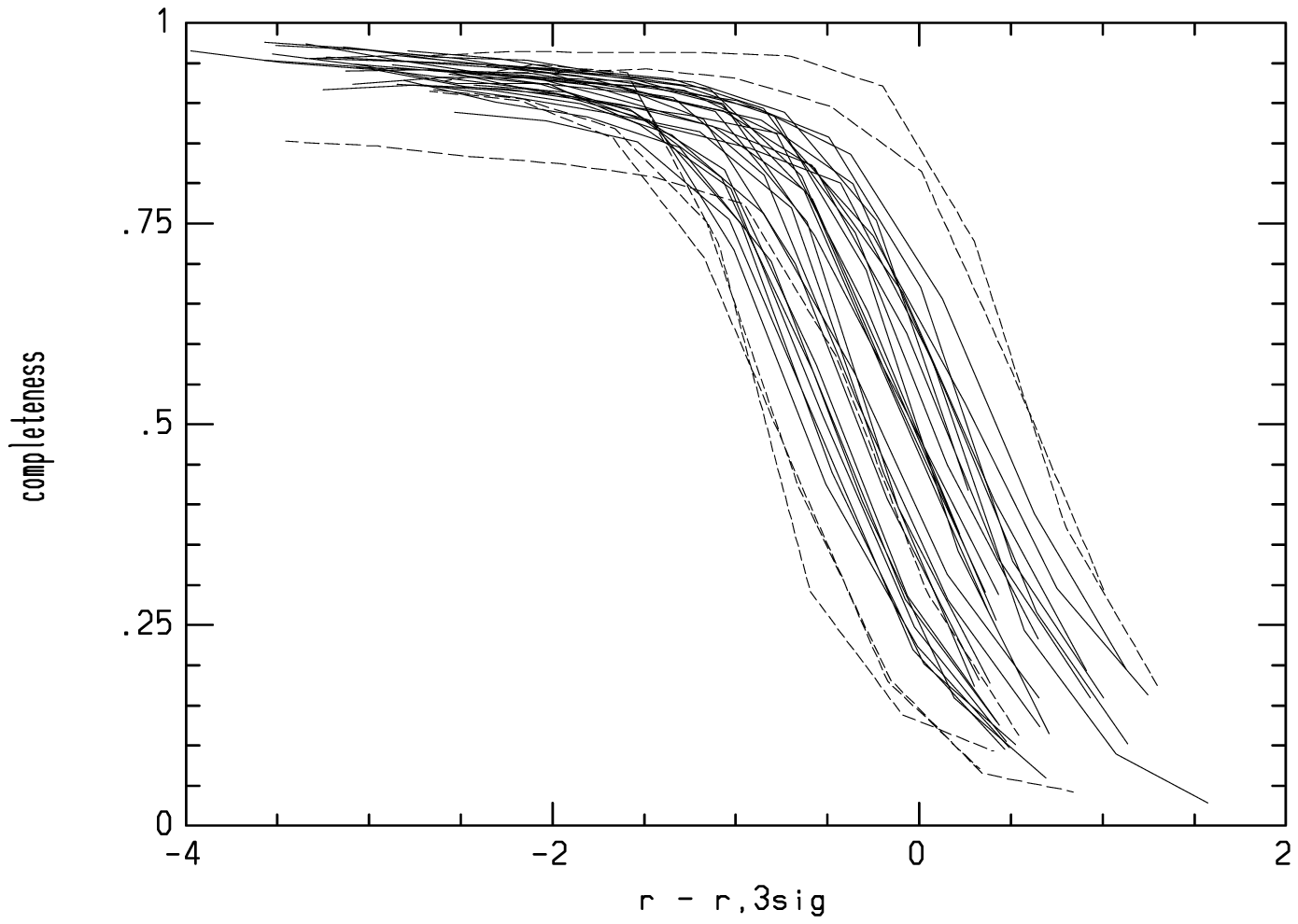}{60pt}{0}{45}{45}{-250}{-153}
\plotfiddle{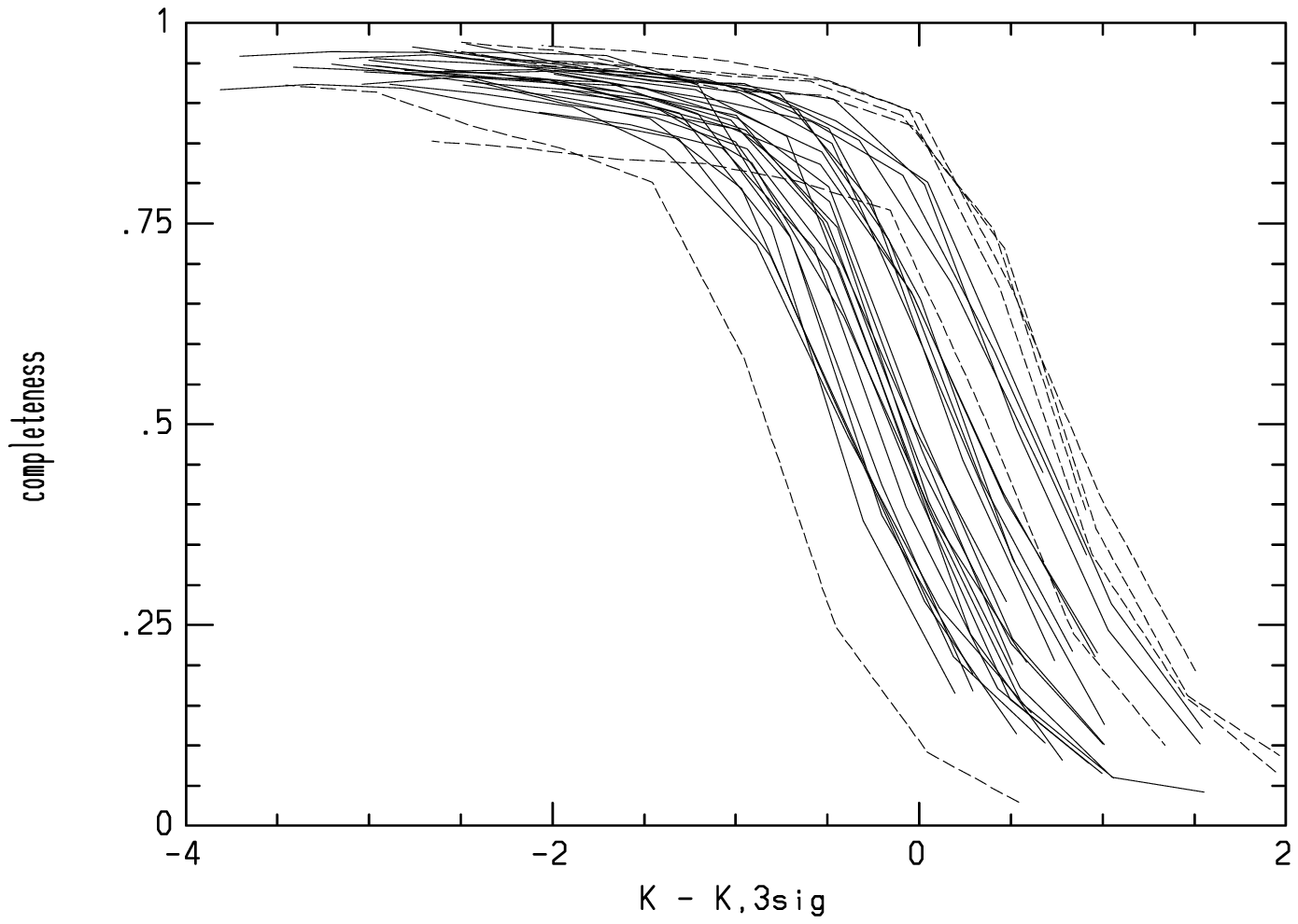}{60pt}{0}{45}{45}{-40}{-80}
\plotfiddle{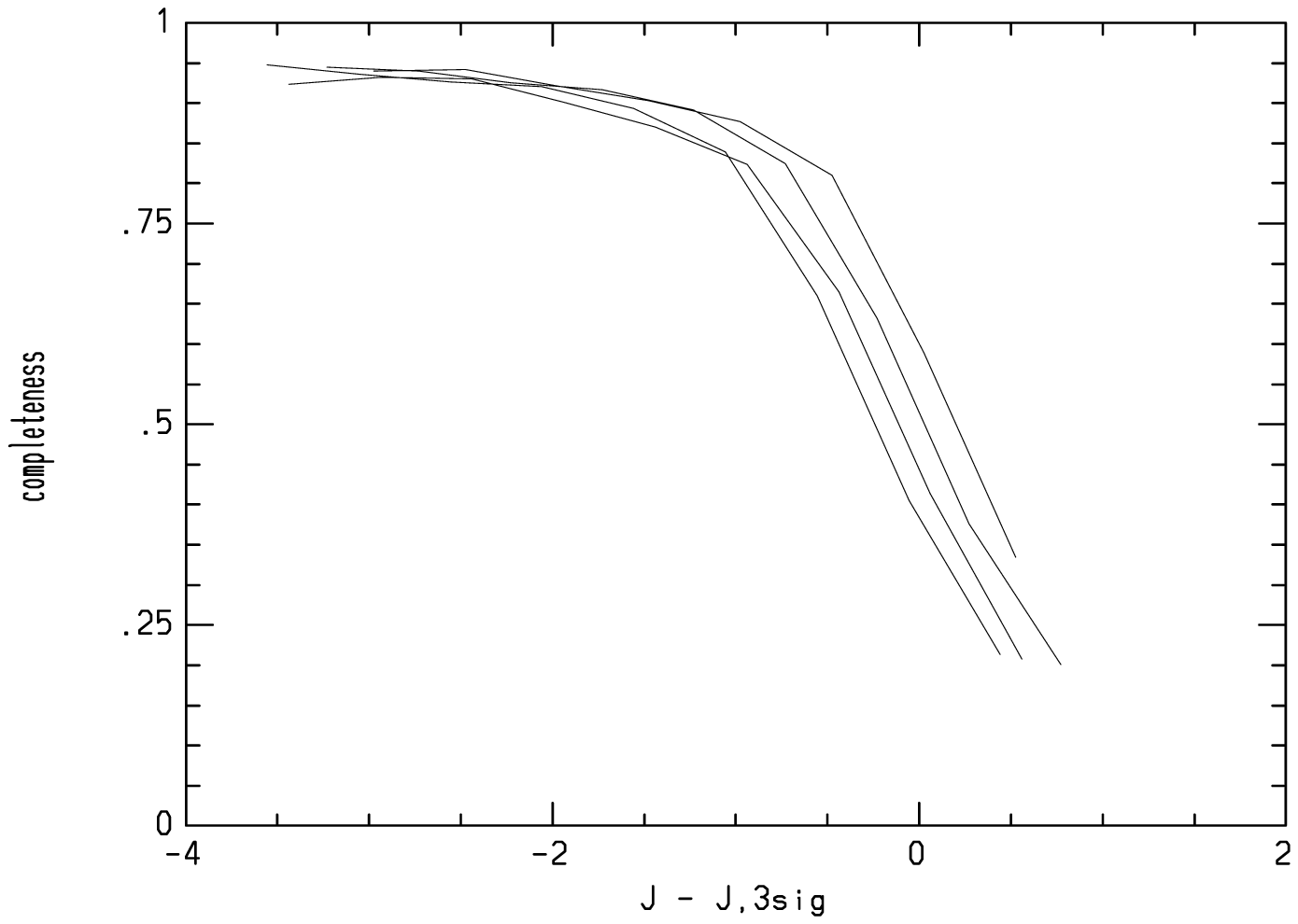}{60pt}{0}{45}{45}{-144}{-130}
\caption[Completeness vs. Magnitude]{
\singlespace
Completeness vs. magnitude relative to 3$\sigma$ limiting magnitude.
a) $r$-band completeness.  b) \ks-band completeness.
c) $J$-band completeness for the four fields where detection was done on 
$r+J+K_s$ images.
Dashed lines indicate fields not used in calculating number-magnitude counts.
All three filters ($rJK_s$) show the expected decline from $>$90\% completeness
at bright magnitudes to $\sim$50\% at the 3$\sigma$ limiting magnitude and
$<$20\% one magnitude fainter than the 3$\sigma$ limit.
}\label{fig_complete}
\end{figure}

The exact procedure followed in adding simulated objects to the images
of each field was as follows.
We took the normalized images from all filters which went into the summed image.
We used FOCAS {\sc clean} to remove from the normalized images objects with 
fewer total counts in the \ks\ filter than an object $\gtrsim$0\fm5 fainter than
the 3$\sigma$ limit in the summed image.  (This step was of course repeated for
brighter counts bins spaced every 0.5 in magnitude.)
We then added unresolved objects 30 at a time 
to the cleaned images in each filter simultaneously, using the PSF of each 
image, a base color of $r-K_s$=4 and $J-K_s$=1.5 for the objects, 
and a random magnitude offset within the magnitude bin in each filter.
(Thus the range of colors simulated is $\pm$0\fm5 around the base values.)
We summed these altered images and reran FOCAS.  
Simulated objects should thus be recovered unless they merged with an object
brighter than themselves in \ks\ and were not split by FOCAS.
(Strictly speaking, this step should be repeated for each filter, each time
removing objects fainter than the simulated objects {\it in that filter}.)
This procedure yields the completeness in all filters simultaneously,
albeit in different magnitude bins given by the average galaxy colors used.
We repeated this procedure 34 times in each counts bin, each time including
different randomly generated Poisson noise appropriate to the object's counts
and the image gain.  We repeated the entire procedure for each field, taking
into account its Galactic extinction.  The results are shown in 
Figure \ref{fig_complete}, plotted against $m-m_{3\sigma}$ to enable direct 
comparison of fields of different limiting magnitudes.
Unusually low completeness is typically due to poor seeing; unusually high
completeness can be due to good seeing or to destriping (see \S\ref{destripe}).
The 50\% completeness magnitudes have a range of $\pm$0\fm5 around the
average 3$\sigma$ limiting magnitude, or equivalently the completeness
at the 3$\sigma$ limiting magnitude has a range of $\pm$30\%.

We use these completeness fractions to correct our observed counts.
Using point sources to determine completeness correction provides
only a lower limit to the true counts of finite-sized galaxies (\cite{mcl95}),
and will bias the results near the detection limit.  However, small average
sizes are found for galaxies at the magnitudes we reach by deep ground-based
imaging (\cite{tys88}) and by $HST$ (\cite{im95}; \cite{roc97}), so point
sources will reasonably approximate the morphology of faint galaxies in the 
$\sim$1.5\arcsec\ seeing of our data.

\subsection{Galaxy Number-Magnitude Relations and Statistical Stellar Contamination Corrections}        \label{SKY}

An obvious first step in comparing our data to previous work 
is to compare the number of galaxies as a function of magnitude,
but prior to that a correction for contamination by stars will be necessary.

Sources of uncertainty in number-magnitude counts include
surface brightness threshold effects,
confusion of multiple objects as a single object,
splitting of a single object into several spurious components,
incompleteness in the galaxy catalog,
inclusion of spurious objects in the galaxy catalog,
errors in the measured galaxy magnitudes,
misclassification of stars as galaxies (and vice versa),
field-to-field variations due to clustering,
and Poisson noise due to the finite number of galaxies detected.
As discusssed in the previous section, the first three effects should be well
accounted for by our completeness corrections,
and the contribution of spurious objects should be small (\cite{hea97}).
Magnitude errors are best 
accounted for by the completeness matrix which we have not calculated, but the 
simple $C(m)$ approach is acceptable.
Poisson noise and field-to-field clustering variations are unavoidable but
their expected strengths can be computed (see \S\ref{angcorr}).

The misclassification of stars as galaxies at faint magnitudes can be
corrected statistically if the observed stellar counts at bright magnitudes
can be extrapolated to fainter magnitudes using a model of the Galaxy.
The Bahcall \& Soneira (1981) Galactic model for $B$ and $V$ has no published 
extensions to other wavelengths, although an estimate of the stellar luminosity
function in $K$ has been made using that model (\cite{ms82}).
Building on the work of Wainscoat {\it et~al.} (1992),
Cohen (1993, 1994, 1995) has developed a Galactic model spanning wavelengths
from 0.14--35~$\mu$m to which we have compared our \ks\ data.
Cohen \& Saracco (1998) and Minezaki \etal\ (1998b) apply the same model to 
their \ks\ field surveys (cf. \cite{min98b}).
This model, dubbed SKY, includes contributions from the disk, bulge, halo,
spiral arms, local spurs, Gould's Belt, and molecular ring of the Galaxy.

We determined the \ks\ magnitude $K_{classlim}$ brighter than which stars were
$robustly$ separated from galaxies by plotting {\tt R} vs. \ks\ for each field.
(Blue stars can be identified to fainter than $K_{classlim}$ by using 
information in the $r$ images, but at $K_s$$<$$K_{classlim}$ all stars should
have been identified, regardless of color.)
We binned in 0\fm5 wide bins the counts of all objects in our fields, of stars
only, and of galaxies only.  The stellar N(m) at \ks$<$$K_{classlim}$ matched
the predictions from SKY to within the observational errors in all fields
except Q~0736$-$063.  The predictions were too high for this field, and we
exclude it from the following analyses.  To obtain the final N(m) for each
field, for each magnitude bin with \ks$<$$K_{classlim}$ we subtracted the
observed stellar counts from the counts of all objects.  
For \ks$>$$K_{classlim}$ we conservatively subtracted the larger of
the observed stellar counts in that bin or the counts as predicted by SKY.
The 1$\sigma$ errors for each magnitude bin were calculated using 
the number of all original objects in the bin, star or galaxy,
and the tables of Gehrels (1986).

We consider briefly the Q~0736$-$063 field.  The Galactic extinction \ebv=0.27
for this field is only an estimate, but this cannot explain the discrepancy.
To bring the observed N(m) in line with predictions would require our $K_s$
magnitudes to be at least 0\fm5 too faint, or \ebv$\geq$1.85.  
The $r$$-$$K_s$ vs. $K_s$ diagram for this field
is if anything biased toward a bluer color distribution than average,
suggesting an overestimated rather than underestimated \ebv,
and unequivocally rules out \ebv$\geq$1.85.
Thus in this field the predicted stellar N(m) was scaled to the observed stellar
N(m) at \ks$<$$K_{classlim}$=16.5 before subtraction from the the galaxy N(m).
The scaling factor was 0.545, or $-$0.263 in the log.  The SKY counts are
estimated to be uncertain by $\sim$0.16 in the log (\cite{min98b}, Figure 2).
The resulting galaxy counts are consistent with those of McLeod \etal\ (1995),
but we do not consider this field in 
future analyses since its \ebv\ is only an estimate.

\begin{figure} 			
\plotone{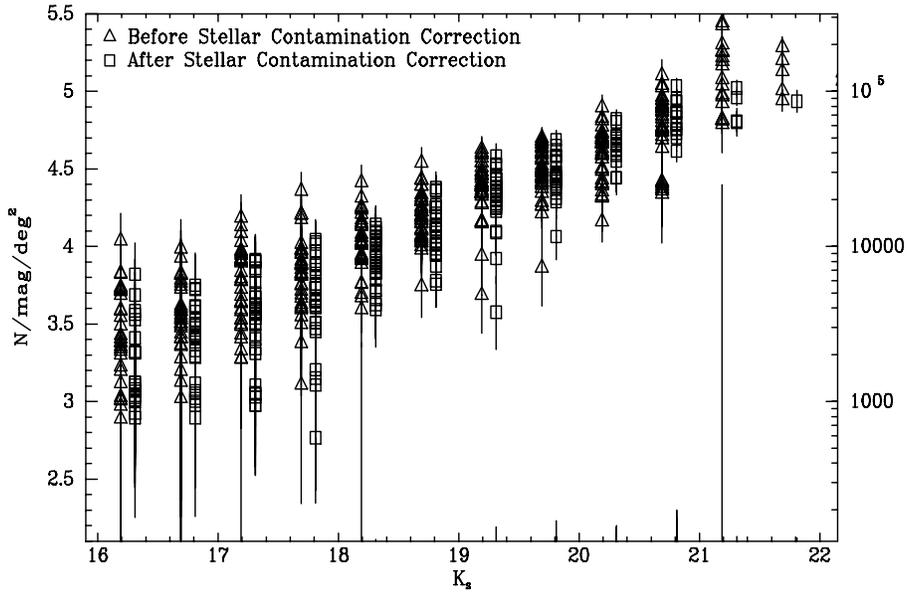}
\caption[$K_s$ N(m) Relation Before and After Stellar Contamination Correction]{
\singlespace
The \ks\ N(m) relation for all 31 good fields with $|b|$$>$20\arcdeg\ is shown
before (triangles) and after (squares) stellar contamination correction, offset
for clarity.  1$\sigma$ Poisson errors are shown for all points.
Post-correction fields are only plotted down to the 50\% completeness magnitude.
}\label{fig_nmb4after1}
\end{figure}

Figure \ref{fig_nmb4after1}
shows the N(m) relations for all our individual fields
before and after correction for stellar contamination as detailed above.
The striking feature of the graph is the large field-to-field scatter at a
given magnitude.  
Post-correction fields are not plotted beyond the 50\% completeness magnitude,
explaining the apparently reduced scatter in those fields at \ks$>$19.
The stellar contamination correction, which noticeably reduces the counts 
at \ks$<$18.5, only slightly reduces the field-to-field scatter.
The field-to-field scatter spans a factor of $\sim$0.4 in the log at \ks=20--21;
i.e. the highest measured surface density per bin is $\sim$2.5 times the lowest.
As shown in \S\ref{angcorr} below, this scatter is consistent with 
expectations for the angular clustering of faint $K$-selected galaxy samples,
which is higher than for faint optically-selected galaxies.

\begin{figure} 	
\epsscale{1.0}
\plotone{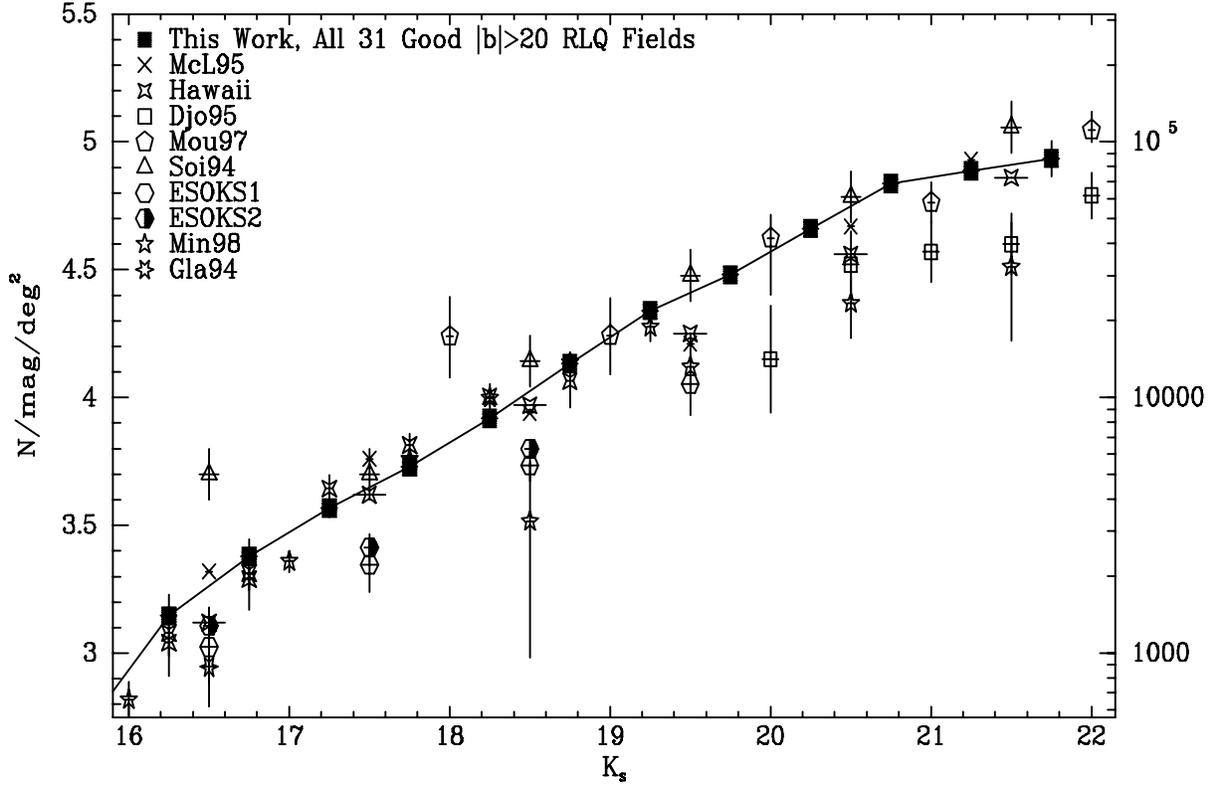}
\caption[$K_s$ band Number-Magnitude Relation vs. Literature]{
\singlespace
The average \ks\ N(m) relation for all 31 good RLQ fields with 
$|b|$$>$20\arcdeg\ is shown as the solid squares connected by the solid line.
1$\sigma$ Poisson errors on N(m) are shown for all our points but are often
smaller than the symbols.  Also shown are 1$\sigma$ RMS uncertainties on the 
zeropoints of the magnitude scales of the different surveys, as given by the 
authors.  (These do not include systematic uncertainties; see \S\ref{syslit}.)
These error bars are also often smaller than the symbols.
Our bins are separated by 0\fm5 magnitudes but have been normalized to number
of galaxies per magnitude bin, as well as per square degree.
N(m) data and 1$\sigma$ Poisson errors from the literature are plotted as 
various symbols.  Uncertainties for the Soifer \etal\ (1994) points are estimates.
Reference codes are as follows:  McL95: McLeod \etal\ (1995); 
Hawaii: Hawaii Medium Deep, Medium Wide, and Deep Surveys: 
Gardner, Cowie \& Wainscoat (1993), Gardner (1995ab), and Cowie \etal\ (1994); 
Gla94: Glazebrook \etal\ (1994); Djo95: Djorgovski \etal\ (1995);
Mou97: Moustakas \etal\ (1997); Soi94: Soifer \etal\ (1994);
ESOKS1 \& ESOKS2: Saracco \etal\ (1997); Min98: Minezaki \etal\ (1998a).
}\label{fig_knm1}
\end{figure}

To determine the average \ks\ N(m) over multiple fields, the stellar
contamination corrected galaxy counts in the desired fields were coadded at
each magnitude, weighting by the area of each contributing field.
We excluded the fields of Q~0736$-$063 (uncertain \ebv)
and Q~1508$-$055 (shallow $K_s$ and no $r$ data),
and magnitude bins with $<$50\% completeness.
The average N(m) for all 31 good RLQ fields with $|b|$$>$20\arcdeg\ (covering
a total area of 221 arcmin$^2$) is shown 
in Figure \ref{fig_knm1} (solid squares + solid line), 
along with counts from the literature (open or half-filled symbols).
Our counts lie within the range found in the literature.
At 16$\leq$\ks$\leq$19, our N(m) agrees quite well with the Hawaii surveys
(\cite{gcw93}; \cite{cow94}) and the survey of McLeod \etal\ (1995).
At \ks$\geq$19, our results are higher than average
and agree most closely with the counts of Soifer \etal\ (1994),
which are as high or higher than any others in the literature, possibly due to
their target fields being around known objects at high redshift.
The excess is large enough 
that it cannot be explained by spurious objects fainter than the 5$\sigma$
limit in each field even if they contaminate the catalog by 10\% (0.04 dex)
at those magnitudes (see \S\ref{catalog}).

\subsubsection{Expected Faint Galaxy Clustering} 	\label{angcorr}

As mentioned in \S\ref{SKY}, the field-to-field variation in galaxy counts
is quite large.  
Poisson errors do not come close to explaining the observed variation.
This result is expected, and is due to galaxy clustering.

We follow Djorgovski \etal\ (1995) in estimating the contribution of 
faint galaxy clustering to the field-to-field variation in our counts 
(see \cite{gla94} for a more exact method).  We assume an
angular correlation function $w(\theta)$=$A_w \theta^{-0.8}$ (all $\theta$
measured in arcsec) and a simple circular top-hat window function of angular
radius $\theta_0$=$(A/\pi)^{1/2}$, where $A$ is the mean area of the fields in 
arcsec$^2$.  The RMS variation due to clustering is then 
\begin{equation}		\label{eq_sigmacl}
1.2 N w(\theta_0)^{1/2}
\end{equation} 		
where $N$ is the mean number of galaxies per field.
Carlberg \etal\ (1997) derive $w(\theta)$=$(1.31\pm0.15)\theta^{-0.8}$
for a sample of $\ge$250 objects with K$\leq$21.5 and $K_{avg}$$\sim$19 
from the Hawaii surveys.  We adopt this relation for our fields.
Note that this clustering amplitude is about a factor of two higher than for 
optically selected galaxies in equivalent magnitude ranges.  This is likely
due to preferential selection of blue galaxies (which are less strongly 
clustered) at optical wavelengths.

\begin{figure} 	
\epsscale{0.725}
\plotone{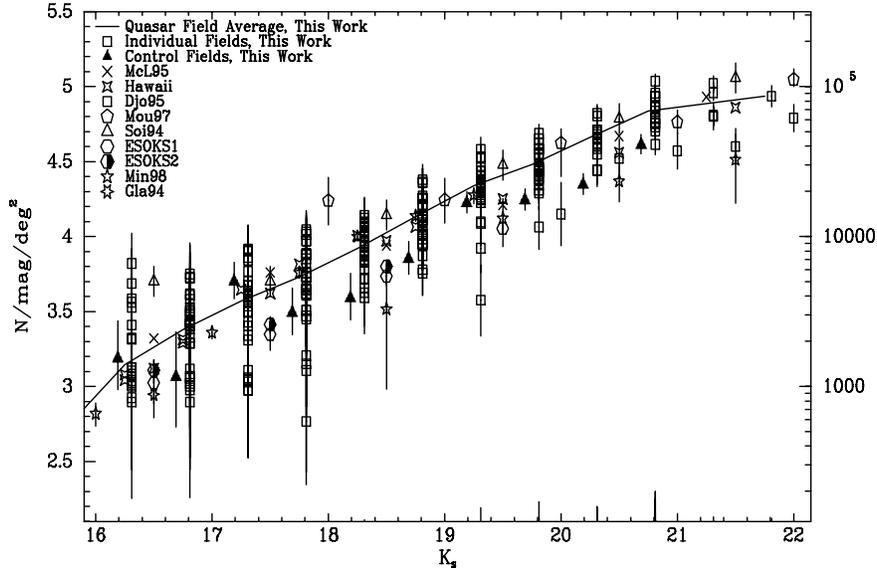}
\caption[$K_s$ N(m) Relation: Individual Fields vs. Literature]{
\singlespace
The average \ks\ N(m) relation for all 31 good fields with
$|b|$$>$20\arcdeg\ is shown as the solid line.
Individual fields are shown down to their 50\% completeness magnitudes
as open squares and 1$\sigma$ Poisson errors (offset +0\fm06 for clarity).
Our control fields (offset $-$0\fm06 for clarity) are shown as filled triangles.
Literature points are the same as in Figure \ref{fig_knm1}.
}\label{fig_b4aftervslit}
\end{figure}

The observed RMS field-to-field variation in our stellar-contamination-corrected
quasar field galaxy counts agrees with the predictions of the Carlberg \etal\ 
$w(\theta)$ to within $\pm$25\% 
in various subsets of the data considered (magnitude limits from \ks=19
to 21, all fields, all 1994 or 1995 fields, all \z$>$1.4 or \z$<$1.4 fields).
Figure \ref{fig_b4aftervslit} shows the counts in our individual fields along 
with points from the literature.  At \ks$<$21, individual field N(m) values
range from less than or equal to the lowest field survey values in the 
literature to above the highest.

\section{Systematic Magnitude Scale Offsets}	\label{systematics}

It is clear from Figure \ref{fig_knm1} that at \ks$\gtrsim$19 our galaxy counts
are higher than those of field surveys; i.e., there is an excess of galaxies in
our combined RLQ fields.  
The larger excess at fainter magnitudes argues that the overall excess is not
simply a random fluctuation in the counts, and in Paper 2 we will show that the
excess does not have the same \rks\ color distribution as the field population
and that at least part of it is spatially concentrated around the quasars.
However, the similarity of the slope of our fields' N(m) counts and the 
literature counts is suggestive of a systematic offset in magnitude scales
being responsible for at least some the difference in the counts.  
Before we can quantify the strength of the galaxy excess
in these fields, we must consider the effects of systematic
errors in our comparison of our N(m) results with the literature.

\subsection{Systematics Between Our Data and the Literature}  \label{syslit}

Three ``$K$'' filters have been used for faint galaxy work: $K$, $K'$
(\cite{wc92}), and $K_s$ (\cite{mcl95}).  These are different passbands by
definition, but the effective passbands can also be significantly altered by 
differences between filter sets at different observatories (see \cite{bb88})
and by the atmospheric transmission at the telescope site.
Fortunately, most faint galaxy studies have been calibrated to one of two
$K$-band photometric systems: the CIT system (\cite{fro78}) to which the Elias
\etal\ (1982) standards are calibrated, or the UKIRT system (\cite{ch92}).
Casali \& Hawarden give the transformations
\begin{eqnarray} \label{eq_kcit}
K_{CIT} = K_{UKIRT} - 0.018 (J-K)_{UKIRT}\\	
(H-K)_{CIT} = 0.960 (H-K)_{UKIRT}
\end{eqnarray}	
Typical faint galaxies with $J$$-$$K$=1.5
should be 0\fm03 fainter on the $K_{UKIRT}$ scale than on the $K_{CIT}$ scale.

Our photometric calibration is to the $K_{UKIRT}$ scale, though we have 
retained the notation \ks\ since our observations were in that filter.  
Soifer \etal\ (1994), Minezaki \etal\ (1998a), and Dickinson \etal\ (1998) also 
observe in \ks\ and calibrate to $K_{UKIRT}$.
Minezaki \etal\ find no color term between \ks\ and $K_{UKIRT}$,
but from the different isophotal wavelengths of the filters they estimate:
\begin{equation}	 \label{eq_ks}
K_s = K_{UKIRT} + 0.04 (H-K)_{UKIRT}    
\end{equation}	
Like Minezaki \etal, we do not make this correction, instead assuming 
\ks=$K_{UKIRT}$ for the UKIRT standards.  If the above relation is correct,
then our \ks=$K_{UKIRT}$+0.02, assuming a typical color of \hk=0.5.
In other words, our $K_s$ magnitude scale (and those of Soifer \etal, 
Minezaki \etal, and Dickinson \etal) may be 0\fm02 fainter 
than the $K_{UKIRT}$ scale to which we claim we are calibrated.
However, that one of us (MC) calculates
$K_s$=$K_{UKIRT}$$-$0\fm019$\pm$0\fm013 for (\hk)$_{UKIRT}$=0, 
so the average difference between $K_s$ and $K_{UKIRT}$ may be less than +0\fm02
for faint galaxies.  We will assume a +0\fm02 difference to be conservative.

Of the $K$ surveys to which we have compared our data, those of McLeod 
\etal\ (1995), EES97, and the Hawaii group (\cite{gcw93}; \cite{cow94}) 
are on the CIT system.
The McLeod \etal\ data were obtained using a $K_s$ filter, and so may require
a color term in the transformation to the CIT system, although they did not
detect one in their data.  Combining equations \ref{eq_kcit} and \ref{eq_ks}
above, with the same assumed average colors, we estimate $K_s$=$K_{CIT}$+0.05.
Most of the remaining surveys, namely Djorgovski \etal\ (1995), 
Glazebrook \etal\ (1994), Moustakas \etal\ (1997), Minezaki \etal\ (1998a),
Soifer \etal\ (1994), and the HDF (\cite{dic98}), are on the UKIRT system,
with the caveat for the latter three discussed in the previous paragraph.  
Lastly, the ESO surveys (Saracco \etal\ 1997) are on the $K'$ system as derived
from observations of Elias standards and the relation
\begin{equation}		 \label{eq_k'}
K' = K_{CIT} + (0.20\pm0.04) (H-K)_{CIT}    
\end{equation}	
given by Wainscoat \& Cowie (1992).  
Combining this with equation \ref{eq_kcit}, we obtain
\begin{equation}		 \label{eq_k''}
K' = K_{UKIRT} - 0.018 (J-K)_{UKIRT} + (0.19\pm0.04) (H-K)_{UKIRT}
\end{equation}	
Assuming average \jk=1.5 and \hk=0.5 as above, the ESO magnitude scale 
can be put on the CIT system by subtracting 0\fm10$\pm$0\fm02
and on the UKIRT system by subtracting 0\fm07$\pm$0\fm02.

These average transformations between systems are accurate to only a few 
percent since faint galaxies will show a range of colors.
The differences between systems are rarely larger than the $\sim$0\fm05 
typical zeropoint uncertainty quoted for the various surveys, although the 
actual uncertainty in the correction to exoatmospheric magnitudes may be larger
if extinction coefficient uncertainties are large.  
Even if the systematic magnitude scale offsets are not large, they may 
be enough to explain the difference between our quasar field counts and the 
literature control field counts.

To test the effects of systematics, we attempt to place all available data on 
the UKIRT $K$ magnitude scale.  (We choose the UKIRT scale because accurate 
flux zeropoints are available for it; see Appendix~\ref{photsys}).
We adjust data from this work, Dickinson \etal, Soifer \etal, and 
Minezaki \etal\ brightwards by 0\fm02.
We adjust the McLeod \etal\ data brightwards by 0\fm02,
and the ESO data (\cite{eso97}) brightwards by 0\fm07.
We adjust all other literature data on the CIT system (namely the Hawaii 
surveys) faintwards by 0\fm03.
We then linearly interpolate the counts back to the original bin centers where
necessary and find the area-weighted average of the N(m) from the different 
surveys.  Since the ESO (\cite{eso97}) and Minezaki \etal\ (1998a) surveys are 
each $\sim$170 arcmin$^2$ in size, they are the 
dominant contributors to the literature control field counts at 17$<$$K$$<$19.

In Figure \ref{fig_nmsysvslit} we plot $N(K_{UKIRT})$ for our data and for the 
area-weighted average of all published random-field imaging surveys which reach
$K_{UKIRT}$$\geq$17 (i.e. excluding Soifer \etal\ 1994, EES97, and Dickinson
\etal\ 1998), including formal uncertainties on the magnitudes.
Area-weighted RMS error bars are plotted for the literature data,
but they probably underestimate the true RMS fluctuations at each
magnitude between fields of size similar to ours ($\sim$8~arcmin$^2$).
since much of the area at $K$$\lesssim$19 comes from large surveys
which are treated as single fields when computing the RMS.
We also plot data from our 1994 and 1995 KPNO 4m runs separately for comparison.
Even with everything calibrated onto the same magnitude scale, there is a
significant offset between our quasar-field data and the control-field data.

Even if our simple offsets have in fact brought all surveys' magnitudes onto
a common scale, it is probable that there are more systematics present in these 
various datasets than we have removed (cf. the discussion of aperture 
correction differences by Djorgovski \etal\ (1995) and the discussion of the
true CIT and UKIRT magnitude scale offsets above).
In particular, near the faint end of all surveys there may be nonlinear
offsets which depend on how ``total'' magnitudes were estimated.
However, the only control fields we have reason to prefer are our own,
and they are too noisy to use by themselves.  Thus our best option is 
to compare our data to the average of all available random-field data
and to remove those systematics we believe we can estimate.

\subsection{Systematics Within Our Data}  \label{syswithin}

\begin{figure} 
\epsscale{0.725}
\plotone{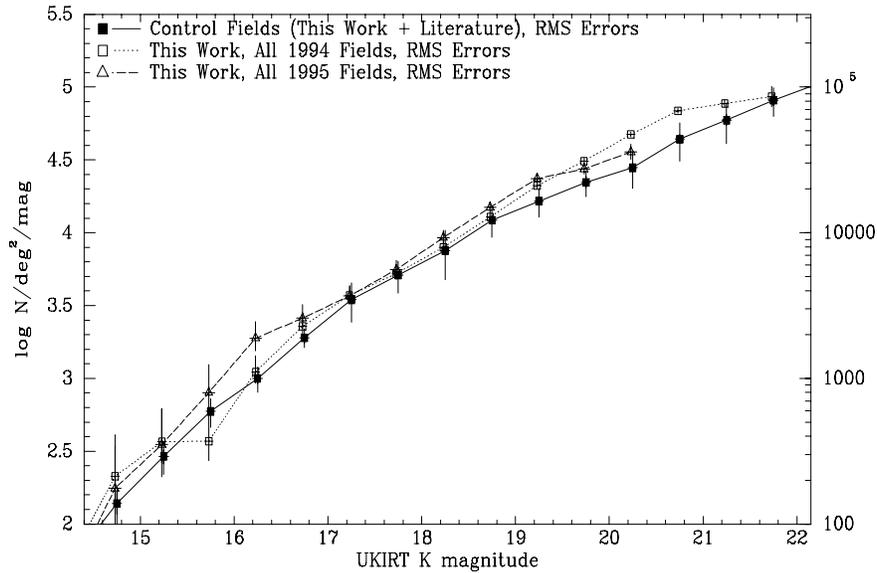}
\caption[$K_{UKIRT}$ N(m) Relation: Our Data vs. Control Field Average]{
\singlespace
The $K_{UKIRT}$ N(m) relations for our 1994 and 1995 KPNO 4m data are 
plotted as dotted and dashed lines respectively.
The area-weighted average of our control fields and all published random-field
imaging surveys (corrected to $K_{UKIRT}$) is plotted as the solid line, along
with RMS errors on the N(m) values and formal uncertainties on the magnitude
bin centers (see text).
}\label{fig_nmsysvslit}
\end{figure}

\begin{figure} 
\plottwo{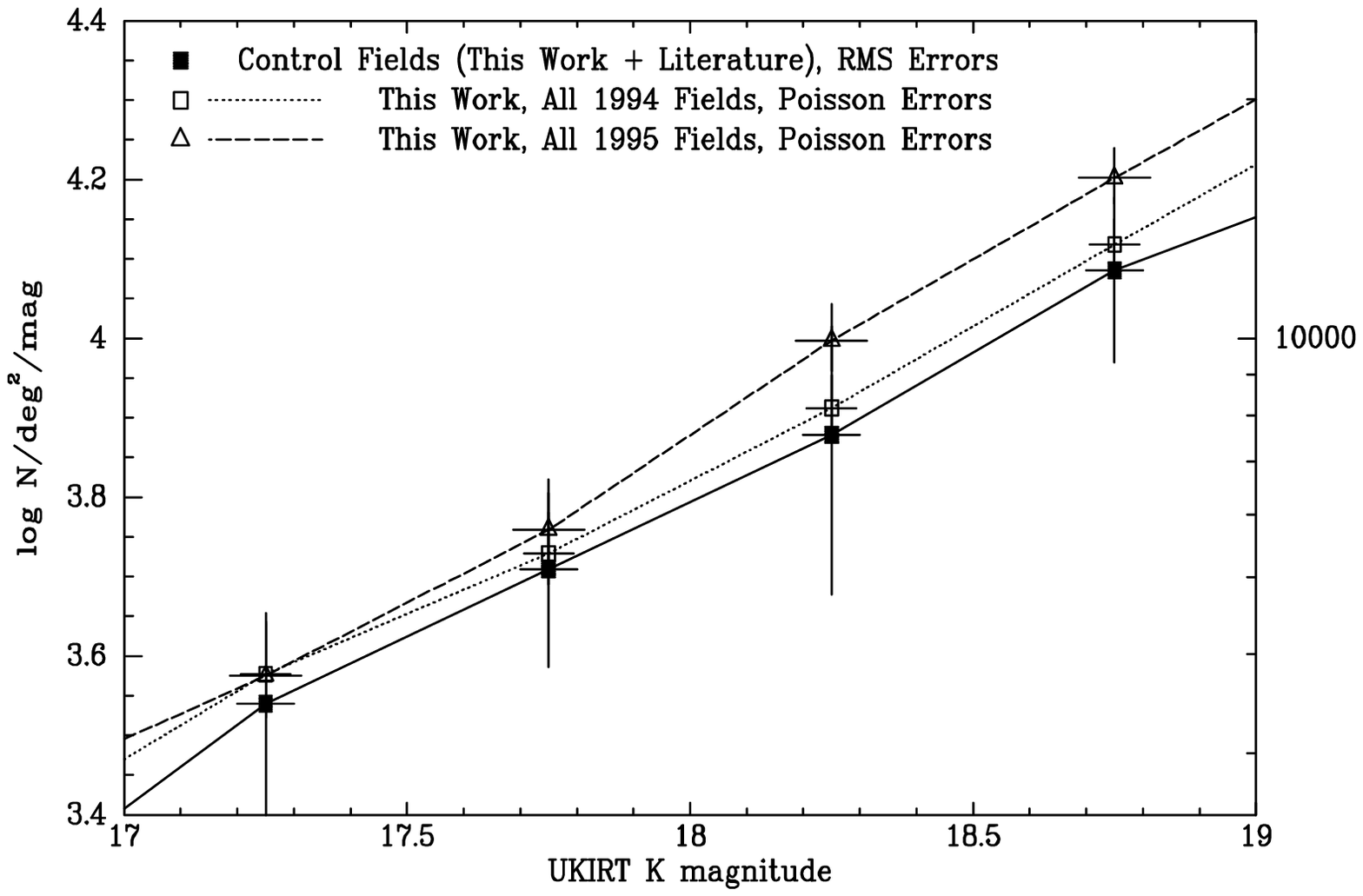}{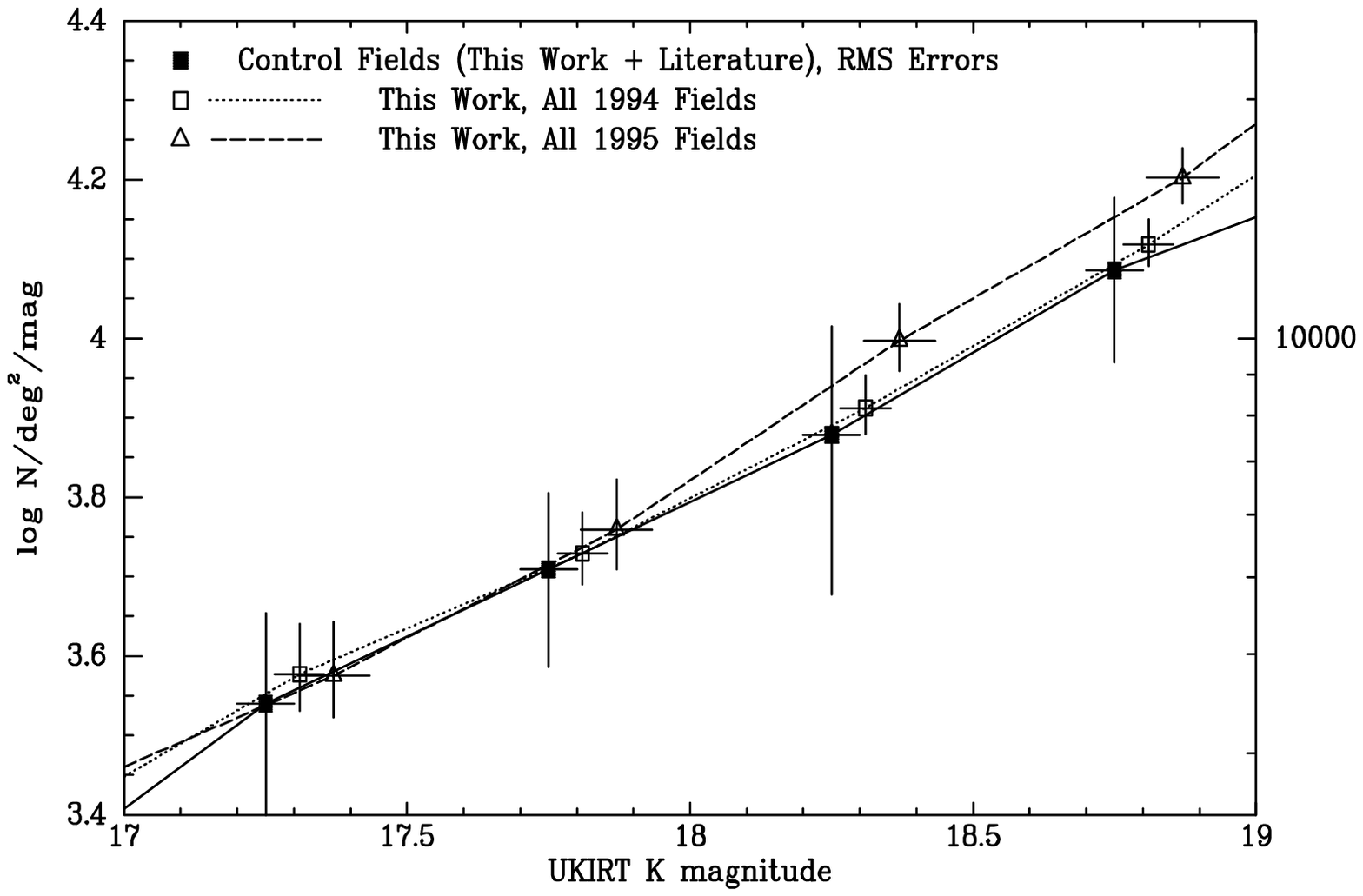}
\caption[N(m) Relation at $K_{UKIRT}$=17--19]{
\singlespace
The $K_{UKIRT}$ N(m) relations for our 1994 and 1995 KPNO 4m run data are
plotted as dotted and dashed lines respectively.
The area-weighted average of our control fields and all published random-field
imaging surveys (corrected to $K_{UKIRT}$) is plotted as the solid line.
RMS errors are plotted for the control field N(m) values, and 
Poisson errors for the quasar fields.
Formal uncertainties are plotted for the magnitude bin centers (see text).
a. The N(m) relation after correction to the UKIRT magnitude scale.
b. The N(m) relation after further correction for systematics in our data.
}\label{fig_nmsysvslit2}
\end{figure}

From Figure \ref{fig_nmsysvslit} we see that in the range \Kuk=17--19 our
1994 and 1995 data agree well in slope, but not in normalization.
Figure \ref{fig_nmsysvslit2}a shows a closeup of this magnitude range.
The RMS errors on the control field data are large enough that essentially all
our data are in agreement with the control field data in this magnitude range,
but the 1995 data lies consistently above the other datasets.
A systematic offset of our 1995 data brightwards of our 1994 data is 
consistent with the possible zeropoint and extinction coefficient systematics
discussed in \S\ref{photkpno}.  

The conservative assumption would be that the offset between our 1994 and 1995
data and between those data and the literature is due to systematic errors.
However, a complication in untangling systematic effects from real differences
is that the 1994 data are composed almost exclusively of \z$>$1.4 quasars, 
and the 1995 data of \z$<$1.4 quasars.
Based on the \Kz\ relation of powerful radio galaxies, 
a brightest cluster galaxy (BCG) might be expected to have 
\Kuk$\sim$17 at \z=1 and \Kuk$\sim$18 at \z=1.4.
Thus it is plausible that the difference between our 1994 and 1995 datasets 
is real, caused by the appearance of cluster galaxies at brighter apparent 
magnitudes in the 1995 (\z$<$1.4) dataset.

\begin{figure} 
\epsscale{0.725}
\plotone{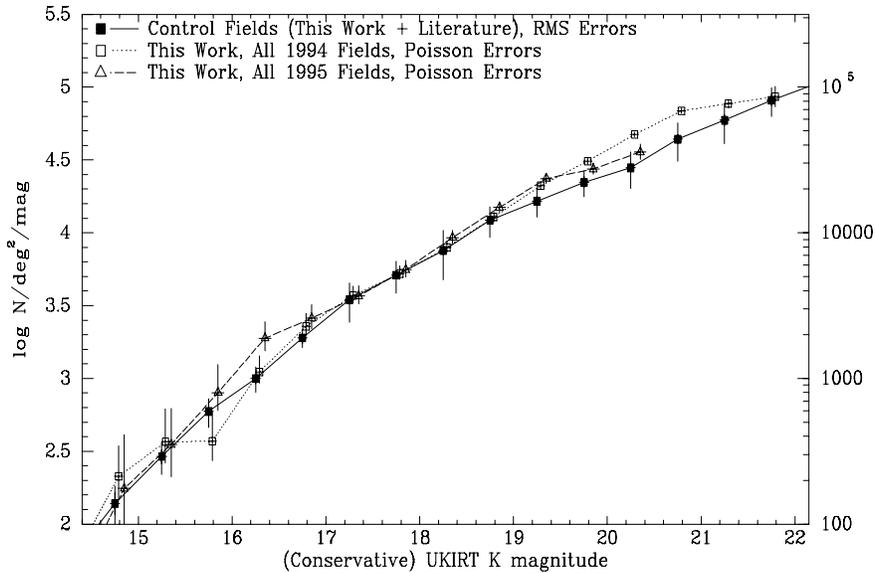}
\caption[Conservative $K_{UKIRT}$ N(m) Relation: Our Data vs. Control Field Average]{
\singlespace
The conservative $K_{UKIRT}$ N(m) relations for our 1994 and 1995 KPNO 4m run 
data are plotted as dotted and dashed lines respectively.
The area-weighted average of our conservative-magnitude control fields and all 
published random-field imaging surveys (corrected to $K_{UKIRT}$) is plotted 
as the solid line.
RMS errors are plotted for the N(m) values and formal uncertainties for the
magnitude bin centers (see text).
}\label{fig_nmsys2vslit}
\end{figure}

Since we cannot $a~priori$ discriminate between these two possibilities, 
we will quantify the galaxy excess for both of them.  
Our liberal assumption is that no further systematics exist once all surveys
have been put on the UKIRT $K$ magnitude scale as described in \S\ref{syslit}.
Thus our $liberal$ magnitudes are $K_{UKIRT}$ magnitudes,
which we estimate are equal to as \ks$-$0\fm02 as discussed in \S\ref{syslit},
and whose N(m) relation is shown in Figure \ref{fig_nmsysvslit}.
Our conservative assumption is that our 1994 and 1995 data should be 
matched to each other and to the control field data at 17$<$\Kuk$<$18; 
i.e. that systematic errors are responsible for the offsets between our two 
datasets and the control fields.
We offset our 1994 $K_{UKIRT}$ magnitudes 0\fm06 faintwards and our 1995
$K_{UKIRT}$ magnitudes 0\fm12 faintwards to form our $conservative$ magnitudes.
As seen in Figure \ref{fig_nmsysvslit2}b, with these offsets our 1994 data
exactly match the control field data at \Kuk=17.5--18, our 1995 data match
at \Kuk=17--17.5, and all three datasets lie within each others' 1$\sigma$
(RMS) error bars at \Kuk=17--18.  Our conservative magnitude scale N(m)
relation is shown in Figure \ref{fig_nmsys2vslit}.

We give our control field counts in Table \ref{tab_knm_cf} and the area-weighted
average of the published literature counts in Table \ref{tab_knm_puball},
both converted to the \Kuk\ system as detailed above.

\section{Conclusions}	\label{conclude1}

We have presented optical and infrared data for a sample of 33 radio-loud
quasars (RLQs) at \z=1--2, discussed our data reduction procedures, and
shown that the galaxy number counts in our fields lie above those of
random-field surveys in the literature.  Paper 2 (Hall \& Green 1998) presents
an analysis and discussion of the excess galaxy population in these fields.

An unusual feature of our data reduction procedure is the image normalization
procedure (\S\ref{coadd}) which allows useful results to be obtained from the
edges of the fields which have less than the full exposure time.  The
complications introduced by this procedure were in the end worth enduring,
since otherwise we would have had very little data at \th$>$80$''$ from the
quasars.  Data at such large angular distances was extremely useful for
confirming the existence and extent of the large-scale excess galaxy population
and for measuring the richnesses of the smaller-scale excesses seen in many
fields (see Paper 2).

The complications faced in establishing the random-field $K$-band counts for
comparison with our quasar-field data 
(\S\ref{syslit}) arose from two sources:  first, the higher field-to-field
variation in $K$-band counts compared to optical counts;
and second, lack of a commonly used well-defined photometric system at $JHK$.
Regarding the first point, it is worth investigating whether random-field $J$
and $H$-band observations show substantially smaller field-to-field variations
than $K$-band observations.  
Regarding the second point, we echo the call of Cohen \etal\ (1992) for the 
adoption of a truly standard system of IR passbands designed to minimize 
site-to-site differences due to atmospheric variations.
One such system is described by Young, Milone \& Stagg (1994).  At the least,
until such time as such a system is widely available, we encourage
extragalactic $JHK$ observers (including ourselves) to pay more attention
to photometric calibration of their data than typically done in the past.
It would also be useful to have standard star measurements in $K_s$ as well as
$K$, since there can be a few percent difference between magnitudes in the two
filters.  Such data should be forthcoming shortly 
from the DENIS (\cite{fou97}) and 2MASS (\cite{skr97}) surveys.  

It is clear from Figure \ref{fig_knm1} that at \ks$\gtrsim$19 there is an excess
of galaxies in our combined RLQ fields.  
The reality of this excess is not in doubt, but its magnitude is
(\S\ref{systematics}).  $K_s$ data from our 1995 KPNO 4m run lies consistently
above the 1994 dataset and the control fields, and a systematic brightwards
offset of our 1995 data is consistent with possible systematic uncertainties in
the $K_s$ zeropoint and extinction coefficient for that run (\S\ref{photkpno}).
Thus we will conduct much of our analysis in Paper 2 in parallel for two
magnitude scales:  conservative (UKIRT $K$ magnitudes, but corrected for each 
dataset so that its bright-end N(m) relation matches the literature control
fields, i.e. assuming the offsets are purely systematic errors) and liberal 
(UKIRT $K$ magnitudes, i.e. assuming no systematics exist beyond those between
our natural $K_s$ and UKIRT $K$ magnitudes as described in \S\ref{syslit}).
Future photometric snapshots of our fields in a $K$ filter on a well-calibrated
system would reduce the uncertainties on the surface density of the excess
galaxy population in these fields.

\acknowledgements

This work was part of a Ph.D. thesis at the University of Arizona.
PBH acknowledges support from an NSF Graduate Fellowship and from NASA.
MC acknowledges the past support of AFRL through Dr. S. D. Price, for
the development of the SKY model under contract F19628-92-C-0090 with
Vanguard Research Inc., and thanks NASA-Ames for partial support
through cooperative agreement NCC 2-142 with Berkeley.
We thank M. Dickinson for providing {\sc dimsum}; D. Minniti, E. Hooper,
C. Liu, B. Golisch, J. Saucedo, and B. Jannuzi for obtaining some of the data
used herein; J. Barnes and F. Valdes for patient IRAF help; the referee for
helpful comments; and Vic Hanson, Dennis Means, 
and the rest of the Steward Observatory telescope operations staff.

This research has made use of observations made at the Kitt Peak National 
Observatory, National Optical Astronomy Observatories, which is operated by 
the Association of Universities for Research in Astronomy, Inc., 
under contract to the National Science Foundation,
and at the Infra-Red Telescope Facility, which is operated by the University
of Hawaii under contract to the National Aeronautics and Space Administration;
the NASA/IPAC Extragalactic Database (NED), operated by the Jet Propulsion
Laboratory, California Institute of Technology, under contract to NASA;
and data from operations made with the NASA/ESA Hubble Space Telescope, 
obtained from the data archive at the Space Telescope Science Institute, 
which is operated by the Association of Universities for Research in Astronomy,
Inc., under NASA contract NAS 5-26555.

\appendix
\section{Photometric Systems and Parameters}  \label{photsys}

In this appendix we tabulate data related to the filters and photometric
systems used in this work.  For our consideration of spectral energy
distributions (Paper 2), for each filter we need to know $S_{\nu}$($m$=0),
the equivalent monochromatic exoatmospheric flux density (usually given in Jy)
at zero magnitude.
It is also useful to know the offset $\Delta$AB to the AB magnitude system,
which by definition has $S_{\nu}$($m$=0)=3631~Jy for all filters (\cite{fuk96}).
Unlike the standard $UBVRI$ photometric system, 
the Thuan-Gunn system calibration is based on BD+17$\arcdeg$4708 
(an F subdwarf) which is defined to have $m=9.5$ in all filters in the system.

We take $S_{\nu}$($m$=0) and $\Delta$AB for $riz$ and $R_TI_C$ from Windhorst
\etal\ (1991; W91), who did their $R_TI_C$ optical imaging at the Steward 
90$''$, and for $R_CI_T$ from Fukugita, Shimasaku \& Ichikawa (1995; FSI95).
The numerical values are given in Table \ref{tab_photsysopt}.  For comparison,
we also tabulate AB offsets from Frei \& Gunn (1994; FG94) and FSI95, and 
calculate $S_{\nu}$($m$=0) from Table 9 of FSI95, taking into account the 
different primary calibration stars for $riz$ and $I$.  Note that FSI95 
refer to our filters as Thuan-Gunn $r$, PFUEI $iz$, and Cousins $RI$.
The largest differences (up to 12.4\%) are for $I_C$.

In this work we deal with near-IR magnitudes calibrated onto both the UKIRT
and CIT scales.  In Table \ref{tab_photsysir} we reproduce the values of
$S_{\nu}$($m$=0) and $\Delta$AB for $JHK$ in both systems from Table 12.2 of
MacKenty \etal\ (1997).  
It is important to remember that it is not just the magnitude scales of the
UKIRT and CIT systems which differ, but the filter passbands as well.  
Thus an object can have $K_{CIT}$=$K_{UKIRT}$ and still have different
equivalent monochromatic exoatmospheric fluxes in the two systems
because the effective wavelengths of the filters are different.
See \S\ref{systematics} for discussion of our conversion between
$K$ and $K_s$ on the UKIRT system.

The most recent published calibration of Vega (\cite{coh92}) indicates that
its $JHK$ fluxes (in Jy) are not equal to the defined zero magnitude fluxes
in either the CIT or UKIRT systems.  
However, since the standards that define the CIT system were calibrated 
assuming $J$=$H$=$K$=0 
(\cite{fro78}; \cite{eli82}) and the UKIRT $K$ zeropoint is defined as 
identical to the CIT system for A0 stars, it can be argued that revised
calibrations of Vega should be propagated through to change the flux 
zeropoints of the CIT and UKIRT systems.  
Thus in Table \ref{tab_photsysir} we reproduce from Tables 1 and 2 of 
Cohen \etal\ (1992) the isophotal wavelengths ($\lambda_{iso}$ = 
$\int$$\lambda F_{vega}(\lambda)S(\lambda)d\lambda$/$\int$$F_{vega}(\lambda)S(\lambda)d\lambda$
where $S(\lambda)$ is the system sensitivity; \cite{coh92}) and the
exoatmospheric monochromatic flux densities at those wavelengths for Vega as
observed in the UKIRT system at Kitt Peak and Mauna Kea.  
We adopt the appropriate one of these $S_{\nu}$($m$=0) values to convert our
UKIRT system magnitudes to exoatmospheric flux densities.

Lastly, we note that the differences in flux zeropoints between magnitude 
scales given in Tables \ref{tab_photsysopt} and \ref{tab_photsysir},
and between other values in the literature not listed here,
are generally small ($\leq$5\%, and at most 10--15\% in a few extreme cases).
However, they will introduce errors of this magnitude into SEDs constructed
from broadband photometry, which will increase the uncertainties and the
frequency of erroneous results when comparing observed and model SEDs.
Other systematic errors are very possibly present in our data
(e.g. extrapolation to airmass=0 and calibration to the UKIRT or CIT scales
regardless of zeropoint), but the addition of another possible source of error 
is still undesirable.

\begin{deluxetable}{lccrccccr}		
\tablecaption{Optical Filter and Photometric System Parameters\label{tab_photsysopt}}
\tablewidth{432.00000pt}
\tablenum{A1}
\tablehead{
\colhead{} & \colhead{F$_{\nu}$(0$^m$),} & \colhead{$\Delta$AB,} & 
\colhead{} &
\colhead{F$_{\nu}$(0$^m$)} & \colhead{$\Delta$AB} & \colhead{$\Delta$AB,} &
\colhead{$\lambda_{eff}$} & \colhead{$\Delta\lambda$} \\[.2ex]
\colhead{Filter} & \colhead{Jy, W91} & \colhead{W91} &
\colhead{R$_{\lambda}$} &
\multicolumn{2}{c}{FSI95/CRL85} & \colhead{FG94} & 
\colhead{(\AA)} & \colhead{(\AA)}}
\startdata
$R_C,R'$ & \nodata & \nodata & 2.0 & 3071 & +0.182 & +0.117 &  6588 & 1568 \\
$R_T$ & 3105 &   +0.176 & 2.0  & 3006 &   +0.205 & +0.055\tablenotemark{a} & 6585 & 1373 \\
$r$ &   4365 & $-$0.194 & 2.19 & 4398 &	$-$0.208 & $-$0.226 &  6677 &  916 \\
$i$ &   4786 & $-$0.294 & 1.60 & 4721 &	$-$0.285 & $-$0.296 &  7973 & 1353 \\
$I_C$ & 2377 &   +0.466 & 1.3  & 2446 &	  +0.429 &   +0.342 &  8060 & 1542 \\
$I_T$ & \nodata & \nodata & 1.3 & 2324 &  +0.484 & +0.309\tablenotemark{a} & 8668 & 1725 \\
$z$ &   4831 & $-$0.304 & 1.20 & 4821 &	$-$0.308 &  \nodata &  9133 &  984 \\
\enddata
\tablenotetext{a}{Highly uncertain value.}
\tablecomments{FG94 is Frei \& Gunn (1994); FSI95 is Fukugita, Shimasaku,
\& Ichikawa (1995); W91 is Windhorst {\it et~al.} (1991).
We have adopted $R_V$=3 (\cite{mb81}) and have renormalized the $R_{\lambda}$
values for Gunn $riz$ from $R_V$=3.05 (\cite{sgh83}).  $R_{\lambda}$ values
for other filters are taken from Mihalas \& Binney (1981).
Zero-magnitude fluxes and AB offsets from various sources are listed in columns
2-3 and 5-7.  
Filter effective wavelengths $\lambda_{eff}$ and widths $\Delta\lambda$ are 
from FSI95.}
\end{deluxetable}

\begin{deluxetable}{cccccccccc}    
\tablecaption{Infrared Filter and Photometric System Parameters\label{tab_photsysir}}
\tablewidth{432.00000pt}
\tablenum{A2}
\tablehead{
\colhead{} & 
\multicolumn{4}{c}{F$_{\nu}$(0$^m$), Jy} &
\colhead{Adopted} & \colhead{} & 
\multicolumn{2}{c}{$\lambda_{iso}$, \micron} &
\colhead{$\Delta\lambda$,} \\[.2ex]
\colhead{Filter} & 
\colhead{Vega@KP} & \colhead{Vega@MK} & \colhead{UKIRT} & \colhead{CIT} & 
\colhead{$\Delta$AB} & \colhead{R$_{\lambda}$} &
\colhead{KP} & \colhead{MK} & 
\colhead{\micron}}
\startdata
$J$      & 1636.6  & 1631.0 & 1600 & 1670 & +0.865\tablenotemark{a} & 0.77 & 1.212 & 1.215 & 0.26 \\
$H$      & 1049.5  & 1049.7 & 1020 & ~980 & +1.347\tablenotemark{a} & 0.50 & 1.654 & 1.654 & 0.29 \\
$K$      & ~653.2  & ~655.0 & ~657 & ~620 & +1.862\tablenotemark{a} & 0.29 & 2.182 & 2.179 & 0.41 \\
$K_s$ & 665$\pm$8 & \nodata & \nodata & \nodata & +1.843\tablenotemark{a} & 0.29 & 2.160 & \nodata & 0.28 \\ 
\enddata
\tablenotetext{a}{Adopted AB offsets are for Mauna Kea for $H$ and for Kitt Peak for all other filters.}
\tablecomments{Exoatmospheric zero-magnitude fluxes are listed in columns 2-3
for UKIRT system filters at Kitt Peak (KP) and Mauna Kea (MK) assuming Vega has
$J$=$H$=$K$=0 (\cite{coh92}),
and in columns 4-5 for the CIT and UKIRT systems (\cite{mac97}).
$R_{\lambda}$ values for all filters are taken from Mihalas \& Binney (1981).
Columns 8-9 list isophotal wavelengths 
from Cohen \etal\ (1992).
For IRIM $K_s$, all quantities listed were calculated by Cohen for this work.}
\end{deluxetable}

\clearpage


\clearpage

\begin{deluxetable}{lrlcrrlcc}    
\tablecolumns{9}
\tablecaption{1$<$$z$$<$2 RLQ Sample: Radio Data\label{tab_rad}}
\scriptsize
\tablenum{1}
\tablehead{
\colhead{} & \colhead{} & \colhead{} &
\colhead{Radio} & \colhead{} & \colhead{} &
\colhead{} & \colhead{$S_{\nu}$, Jy} & \colhead{$P_{rad}$, W/Hz} \\[.2ex]
\colhead{} & \colhead{} & \colhead{$\alpha_r$} &
\colhead{Morph.} & \colhead{LAS} & \colhead{LLS} &
\colhead{LAS} &  \colhead{(5~GHz,} & \colhead{(5~GHz,} \\[.2ex]
\colhead{Name} & \colhead{$\alpha_r$} & \colhead{Refs} &
\colhead{Class} & \colhead{($''$)} & \colhead{(kpc)} &
\colhead{Refs} & \colhead{obs.)} & \colhead{rest)}}
\startdata
\cutinhead{$z$$<$1.4 Subsample}
0003$-$003   & 0.84 & Stickel         & FRII   &   5.50 &  37.92 & mvk91      &  1.400 & 27.72 \nl
0149+218   & 0.20 & 2.7/5.0      & CL     &   4.50 &  32.65 & mbp93      &  1.080 & 27.63 \nl
1328+254   & 0.59 & Stickel         & C      &   0.20 &   1.39 & huo83      &  3.240 & 28.02 \nl
1430-0046 & 1.16 & 1.4/8.4\tablenotemark{a} & \nodata & \nodata &   \nodata & \nodata     &  0.027\tablenotemark{b} & 26.09\tablenotemark{b} \nl
1437+624 & 0.78 & 1.4/5.0      & C      & $<$0.13 & $<$0.91 & vmh84      &  0.796 & 27.51 \nl
1508$-$055   & 0.30 & Stickel         & C      &   3.00 &  21.35 & huo83      &  2.330 & 27.90 \nl
1606+106   & -0.42 & Stickel         & CL     &   6.50 &  46.53 & mbp93      &  1.700 & 27.54 \nl
1718+481 & \nodata & \nodata    & C/CE?  & $<$15.00 & $<$104.56 & kel94      &  0.124 & 26.60\tablenotemark{c} \nl
1739+522   & -0.68 & Stickel         & C      &   1.00 &   7.31 & nhg89      &  1.130 & 27.36 \nl
2144+092   & -0.23 & Stickel         & CE     &   2.30 &  16.13 & mbp93      &  1.010 & 27.30 \nl
2230+114   & 0.50 & Stickel         & T      &   2.40 &  16.55 & nhg89      &  3.650 & 28.03 \nl
2325+293   & 0.97 & 1.4/5.0      & T      &  50.40 & 345.62 & huo83      &  0.428 & 27.22 \nl
\cutinhead{$z$$>$1.4 Subsample}
0017+154   &  1.20 & 2.7/5.0         & T      &  14.00 & 105.84 & bri94      &  0.500 & 28.17 \nl
0033+098   &  0.54 & 1.4/5.0         & T      &   8.00 &  60.38 & law86      &  0.330 & 27.63 \nl
0232$-$042   &  0.49 & 2.7/5.0         & T      &  13.10 &  96.30 & huo83      &  0.620 & 27.58 \nl
0256$-$005   &  0.26 & 2.7/5.0         & C      & $<$1.00 & $<$7.56 & laf94      &  0.230 & 27.38 \nl
0352+123   &  0.63 & 2.7/5.0         & T      &   7.00 &  52.16 & nhg89      &  0.270 & 27.39 \nl
0736$-$063   &  0.24 & 2.7/5.0         & CL     &   1.40 &  10.56 & nhg89      &  1.190 & 28.04 \nl
0808+289   &  0.69 & 1.4/5.0         & T      &  34.00 & 256.41 & bar88      &  0.053 & 26.89 \nl
0831+101   &  0.80 & 1.4/5.0         & T      &  30.00 & 225.30 & nhg89      &  0.074 & 27.00 \nl
0835+580   &  0.86 & 1.4/5.0         & FRII   &  16.00 & 118.60 & mvk91      &  0.688 & 27.84 \nl
0926+117   &  0.60 & 2.7/5.0         & T      &   7.00 &  52.55 & bar88      &  0.180 & 27.29 \nl
0952+179   &  0.39 & 2.7/5.0         & C?     & $\lesssim$2.00 & $\lesssim$14.75 & swa86      &  0.740 & 27.64 \nl
1018+348   &  0.01 & 1.4/5.0         & T      &  19.00 & 139.20 & mc83       &  0.469 & 27.25 \nl
1126+101   &  0.05 & 2.7/5.0         & T      &  19.00 & 140.64 & bar88      &  0.310 & 27.16 \nl
1218+339   &  0.90 & 1.4/5.0         & FRII   &  10.00 &  74.05 & mvk91      &  0.869 & 27.95 \nl
1221+113   &  0.71 & 1.4/5.0         & T?     &   2.00 &  15.02 & bar88      &  0.146 & 27.26 \nl
1258+404   &  1.14 & 1.4/5.0         & FRII   &  23.00 & 171.90 & mvk91      &  0.301 & 27.69 \nl
1416+067   &  0.96 & Stickel         & T      &   1.60 &  11.75 & swa86      &  1.500 & 28.14 \nl
1556+335   &  0.33 & 1.4/5.0         & C      & $<$1.00 & $<$7.47 & swa86      &  0.870 & 27.80 \nl
2044$-$168   &  0.06 & 2.7/5.0         & CL     &  12.00 &  90.61 & nhg89      &  0.800 & 27.80 \nl
2149+212   &  0.60 & 2.7/5.0         & T?     &   2.00 &  14.83 & lbm93      &  0.360 & 27.46 \nl
2345+061   &  0.83 & 2.7/5.0         & T?     &   1.10 &   8.16 & lbm93      &  0.270 & 27.43 \nl
\enddata
\tablecomments{Radio 
spectral $\alpha_r$ is for $S_{\nu}$ $\propto$ $\nu^{-\alpha_r}$.
References for $\alpha_r$ are Stickel (personal communication)
or NED if two numbers are given; the numbers are the observed frequencies 
(in GHz) between which $\alpha_r$ was calculated.  All $\alpha_r$ provided 
by Stickel were measured between 2.7 and 5~GHz, observed.
We have made no attempt to correct for radio variability,
and since much of the data was obtained at different epochs our values of
$\alpha_r$ should be considered representative only.
Radio morphological categories are primarily from Neff \& Hutchings (1990): 
C = core only; CE = extended core; CL = core + lobe; and T = core + 2-sided
lobe (triple).  We also include FRII = 2 edge-brightened lobes, with or without
a core (McCarthy, van Breugel \& Kapahi 1991).  There is some overlap between
the FRII and T categories since different references do not always distinguish
between them.
LLS (Largest Linear Size) calculated for $h$=0.75, $q_0$=0.1.
LAS (Largest Angular Size) reference codes given below.
}
\tablenotetext{a}{8.4~GHz flux from Visnovsky {\it et al.} 1992.}
\tablenotetext{b}{$S_{\nu}$ and $P_{rad}$ calculated from 8.4~GHz observations and 1.4/8.4~GHz spectral slope.}
\tablenotetext{c}{$P_{rad}$ calculated assuming $\alpha_r$=0.5.}
\tablerefs{au85: Antonucci \& Ulvestad 1985;
bar88: Barthel {\it et~al.} 1988;
bp86: Browne \& Perley 1986;
bri94: Bridle {\it et~al.}\ 1994;
hpg88: Hutchings, Price \& Gower 1988;
huo83: Hintzen, Ulvestad \& Owen 1983;
hut96: Hutchings {\it et~al.} 1996;
kel94: Kellermann {\it et~al.} 1994;
laf94: La Franca {\it et~al.} 1994;
law86: Lawrence {\it et~al.} 1986;
lbm93: Lonsdale, Barthel \& Miley 1993;
mbp93: Murphy, Browne \& Perley 1993;
mc83: Machalski \& Condon 1983;
mvk91: McCarthy, van Breugel \& Kapahi 1991;
nh90: Neff \& Hutchings 1990;
nhg89: Neff, Hutchings \& Gower 1989;
nil93: Nilsson {\it et~al.} 1993;
pri93: Price {\it et~al.} 1993;
rse95: Rector, Stocke \& Ellingson 1995;
swa86: Swarup {\it et~al.} 1986;
wb86: Wills \& Browne 1986;
wil79: Wills 1979;
vmh84: van Breugel, Miley \& Heckman 1984.
}
\end{deluxetable}

\begin{deluxetable}{llccrrlcc}    
\tablecolumns{9}
\tablecaption{1$<$$z$$<$2 RLQ Sample: Basic Data\label{tab_zg1}}
\footnotesize
\tablenum{2}
\tablehead{
\colhead{Name} & \colhead{Alt.} &
\colhead{RA} & \colhead{Dec} &
\colhead{} & \colhead{} &
\colhead{} & \colhead{} & \colhead{Ass.} \\[.2ex]
\colhead{(1950)} & \colhead{Name} &
\colhead{(1950)} & \colhead{(1950)} &
\colhead{l} & \colhead{b} &
\colhead{$z$} & \colhead{$\alpha_r$} & \colhead{Abs.}}
\startdata
\cutinhead{$z$$<$1.4 Subsample}
0003$-$003   &   3C2   &   00:03:48.8   &   $-$00:21:07   &   99.2803   &   $-$60.8592   &   1.037   &   S   &   ?    \nl
0149+218   &   PKS   &   01:49:31.8   &   +21:52:21   &   141.0586   &   $-$38.6010   &   1.320   &   F   &   ?    \nl
1328+254   &   3C287   &   13:28:15.9   &   +25:24:37   &   22.4665   &   +80.9884   &   1.055   &   S   &   N    \nl
1430-0046   &   LBQS   &   14:30:09.9   &   $-$00:46:04   &   347.9486   &   +52.7950   &   1.0229   &   S   &   ?    \nl
1437+624   &   OQ663   &   14:37:32.0   &   +62:24:48   &   103.5246   &   +50.6938   &   1.090   &   S   &   N    \nl
1508$-$055   &   4C-05.64   &   15:08:14.9   &   $-$05:31:49   &   353.9090   &   +42.9358   &   1.191   &   F   &   N    \nl
1606+106   &   4C10.45   &   16:06:23.4   &   +10:37:00   &   23.0283   &   +40.7886   &   1.226   &   F   &   ?    \nl
1718+481   &   PG   &   17:18:17.6   &   +48:07:11   &   74.3750   &   +34.8295   &   1.084   &   ?   &   ?    \nl
1739+522   &   4C51.37   &   17:39:29.0   &   +52:13:10   &   79.5635   &   +31.7482   &   1.379   &   F   &   ?    \nl
2144+092   &   PKS   &   21:44:42.5   &   +09:15:51   &   65.7957   &   $-$32.2623   &   1.113   &   F   &   ?    \nl
2230+114   &   4C11.69   &   22:30:07.8   &   +11:28:23   &   77.4379   &   $-$38.5824   &   1.037   &   S   &   Y    \nl
2325+293   &   4C29.68   &   23:25:42.2   &   +29:20:39   &   102.0649   &   $-$29.8561   &   1.015   &   S   &   ?    \nl
\cutinhead{$z$$>$1.4 Subsample and Control Fields}
0017+154   &   3C9   &   00:17:49.8   &   +15:24:16   &   112.0466   &   $-$46.5332   &   2.012   &   S   &   N    \nl
0033+098   &   4C09.01   &   00:33:48.2   &   +09:51:29   &   116.8355   &   $-$52.5604   &   1.920   &   S   &   Y    \nl
0232$-$042   &   4C-04.06   &   02:32:36.6   &   $-$04:15:11   &   174.4627   &   $-$56.1557   &   1.438   &   F   &   Y    \nl
0256$-$005   &   PKS   &   02:56:55.1   &   $-$00:31:55   &   177.1898   &   $-$49.2283   &   1.995   &   F   &   Y    \nl
0352+123   &   4C12.17   &   03:52:59.2   &   +12:23:03   &   177.4171   &   $-$30.2463   &   1.608   &   S   &   Y    \nl
0736$-$063   &   PKS   &   07:36:30.2   &   $-$06:20:03   &   224.1712   &   +7.5168   &   1.901   &   F   &   Y    \nl
0808+289   &   B2   &   08:08:32.1   &   +28:54:02   &   193.4684   &   +29.1280   &   1.887   &   S   &   Y    \nl
0831+101   &   87GB   &   08:31:57.6   &   +10:08:17   &   215.6129   &   +27.4227   &   1.760   &   S   &   Y    \nl
0835+580   &   3C205   &   08:35:10.0   &   +58:04:52   &   159.2600   &   +36.8963   &   1.534   &   S   &   Y    \nl
0926+117   &   4C11.32   &   09:26:01.0   &   +11:47:32   &   220.7837   &   +40.0883   &   1.750   &   S   &   N    \nl
0952+179   &   PKS   &   09:52:11.8   &   +17:57:45   &   216.4550   &   +48.3636   &   1.472   &   F   &   N    \nl
1018+348   &   B2   &   10:18:24.1   &   +34:52:29   &   190.3893   &   +57.1220   &   1.404   &   F   &   Y    \nl
1126+101   &   PKS   &   11:26:38.7   &   +10:08:32   &   250.7199   &   +64.0478   &   1.516   &   F   &   Y    \nl
1218+339   &   3C270.1   &   12:18:03.9   &   +33:59:50   &   166.3079   &   +80.6390   &   1.520   &   S   &   Y    \nl
1221+113   &   MC2   &   12:21:47.4   &   +11:24:00   &   279.9186   &   +72.7723   &   1.755   &   S   &   N    \nl
1258+404   &   3C280.1   &   12:58:14.1   &   +40:25:15   &   115.2582   &   +76.8404   &   1.660   &   S   &   N    \nl
1416+067   &   3CR298   &   14:16:38.8   &   +06:42:21   &   352.1603   &   +60.6665   &   1.430   &   S   &   Y    \nl
1556+335   &   87GB   &   15:56:59.4   &   +33:31:47   &   53.5683   &   +49.3725   &   1.646   &   F   &   Y    \nl
2044$-$168   &   PKS   &   20:44:30.8   &   $-$16:50:09   &   29.9668   &   $-$32.9346   &   1.937   &   F   &   Y    \nl
2149+212   &   4C21.59   &   21:49:26.1   &   +21:16:07   &   76.5797   &   $-$24.7840   &   1.536   &   S   &   N    \nl
2345+061   &   4C06.76   &   23:45:58.4   &   +06:08:19   &   96.2347   &   $-$53.1630   &   1.540   &   S   &   N    \nl
CtrlFld1   &   \nodata   &   00:59:09.3   &   $-$00:59:55   &  128.6937   &   $-$63.4866   &   \nodata & \nodata & \nodata \nl
CtrlFld3   &   \nodata   &   00:59:39.7   &   $-$00:51:18   &  128.9466   &   $-$63.3321   &   \nodata & \nodata & \nodata \nl
\enddata
\tablecomments{``Name'' is the quasar's coordinate designation in epoch B1950.
$\alpha_r$ refers to radio spectral slope, given as S
(steep spectrum), F (flat spectrum), ? (unknown), or Q (radio-quiet quasar).
See Table 1 
for detailed radio properties.
``Ass. Abs.'' refers to the presence or absence of associated \ion{C}{4} and/or
\ion{Mg}{2} absorption in the quasars' spectra; an ? indicates that no
information is available.  See Table 3 
for detailed intervening and associated absorption information.}
\end{deluxetable}

\begin{deluxetable}{llcccclc}    
\tablecolumns{8}
\tablecaption{1$<$$z$$<$2 RLQ Sample: Absorption Line Data\label{tab_abs}}
\scriptsize
\tablenum{3}
\tablehead{
\colhead{} & \colhead{} &
\colhead{} & \colhead{} &
\colhead{Redshift} & \colhead{} &
\colhead{} & \colhead{Ass.} \\[.2ex]
\colhead{Name} & \colhead{$z_{quasar}$} &
\colhead{$z_{abs}$} & \colhead{Ion} &
\colhead{Search Interval} & \colhead{Ion} &
\colhead{References} & \colhead{Abs.?}}
\startdata
\cutinhead{$z$$<$1.4 Subsample}
0003$-$003   &   1.037   &   \nodata   &   \phm{x}   &   \nodata   &   \phm{x}   &   \phm{x}   &   \phm{x}   \nl
0149+218   &   1.320   &   \nodata   &   \phm{x}   &   \nodata   &   \phm{x}   &   \phm{x}   &   \phm{x}   \nl
1328+254   &   1.055   &   none   &   \phm{x}   &   0.59-1.06   &   \ion{Mg}{2}   &   abe94   &   \phm{x}   \nl
1430-0046   &   1.0229   &   \nodata   &   \phm{x}   &   \nodata   &   \phm{x}   &   \phm{x}   &   \phm{x}   \nl
1437+624   &   1.090   &   0.8723   &   \ion{Mg}{2}   &   0.63-1.09   &   \ion{Mg}{2}   &   abe94   &   \phm{x}   \nl
1508$-$055   &   1.191   &   none   &   \phm{x}   &   1.02-1.20   &   \ion{C}{4}   &   abe94   &   \phm{x}   \nl
\phm{x}   &   \phm{x}   &   \phm{x}   &   \phm{x}   &   0.12-1.20   &   \ion{Mg}{2}   &   abe94   &   \phm{x}   \nl
1606+106   &   1.226   &   \nodata   &   \phm{x}   &   \nodata   &   \phm{x}   &   \phm{x}   &   \phm{x}   \nl
1718+481   &   1.084   &   \nodata   &   \phm{x}   &   \nodata   &   \phm{x}   &   \phm{x}   &   \phm{x}   \nl
1739+522   &   1.379   &   \nodata   &   \phm{x}   &   \nodata   &   \phm{x}   &   \phm{x}   &   \phm{x}   \nl
2144+092   &   1.113   &   \nodata   &   \phm{x}   &   \nodata   &   \phm{x}   &   \phm{x}   &   \phm{x}   \nl
2230+114   &   1.037   &   $\sim$1.037   &   \ion{C}{4}   &   1.00-1.05   &   \ion{C}{4}   &   wil95   &   Y   \nl
\phm{x}   &   \phm{x}   &   \phm{x}   &   \phm{x}   &   0.13-1.04   &   \ion{Mg}{2}   &   abe94   &   \phm{x}   \nl
2325+293   &   1.015   &   \nodata   &   \phm{x}   &   \nodata   &   \phm{x}   &   \phm{x}   &   \phm{x}   \nl
\cutinhead{$z$$>$1.4 Subsample}
0017+154   &   2.012   &   1.3636   &   \ion{Mg}{2},\ion{C}{4}   &   1.15-2.25   &   \ion{C}{4}   &   y91   &   \phm{x}   \nl
\phm{x}   &   \phm{x}   &   1.6250   &   \ion{Mg}{2},\ion{C}{4}   &   0.2-0.85   &   \ion{Mg}{2}   &   y91   &   \phm{x}   \nl
\phm{x}   &   \phm{x}   &   1.8723   &   \ion{C}{4}   &   1.1-1.25   &   \ion{Mg}{2}   &   y91   &   \phm{x}   \nl
\phm{x}   &   \phm{x}   &   1.9382   &   \ion{C}{4}   &   1.48-1.67   &   \ion{Mg}{2}   &   jhb91   &   \phm{x}   \nl
0033+098   &   1.92   &   1.7776   &   \ion{Mg}{2},\ion{C}{4}   &   1.46-1.95   &   \ion{C}{4}   &   ss92,fun   &   \phm{x}   \nl
\phm{x}   &   \phm{x}   &   1.9036   &   \ion{Mg}{2},\ion{C}{4}   &   0.85-1.95   &   \ion{Mg}{2}   &   ss92,fun   &   Y   \nl
0232$-$042   &   1.438   &   1.425   &   \ion{C}{4}   &   1.0-1.45   &   \ion{C}{4}   &   y91   &   Y   \nl
\phm{x}   &   \phm{x}   &   \phm{x}   &   \phm{x}   &   0.1-1.45   &   \ion{Mg}{2}   &   \phm{x}   &   \phm{x}   \nl
0256$-$005   &   1.9951   &   $\sim$1.9951   &   \ion{C}{4}   &   1.58-2.05   &   \ion{C}{4}   &   fun   &   Y   \nl
\phm{x}   &   \phm{x}   &   \phm{x}   &   \phm{x}   &   0.43-0.75   &   \ion{Mg}{2}   &   fun   &   Y   \nl
0352+123   &   1.60   &   1.4831   &   \ion{Mg}{2}   &   1.55-1.62   &   \ion{C}{4}   &   jhb91   &   \phm{x}   \nl
\phm{x}   &   \phm{x}   &   1.5971   &   \ion{C}{4}   &   0.4-1.7   &   \ion{Mg}{2}   &   btt90   &   Y   \nl
\phm{x}   &   \phm{x}   &   1.6007   &   \ion{Mg}{2}   &   \phm{x}   &   \phm{x}   &   btt90   &   Y   \nl
0736$-$063   &   1.901   &   1.2009   &   \ion{Mg}{2}   &   1.25-1.95   &   \ion{C}{4}   &   abe94   &   \phm{x}   \nl
\phm{x}   &   \phm{x}   &   1.2035   &   \ion{Mg}{2}   &   0.25-1.91   &   \ion{Mg}{2}   &   abe94   &   \phm{x}   \nl
\phm{x}   &   \phm{x}   &   1.8175   &   \ion{C}{4}   &   \phm{x}   &   \phm{x}   &   abe94   &   \phm{x}   \nl
\phm{x}   &   \phm{x}   &   1.9131   &   \ion{C}{4}   &   \phm{x}   &   \phm{x}   &   y91   &   Y   \nl
\phm{x}   &   \phm{x}   &   1.9310   &   \ion{C}{4}   &   \phm{x}   &   \phm{x}   &   y91   &   Y   \nl
0808+289   &   1.887   &   0.6492   &   \ion{Mg}{2}   &   1.65-1.9   &   \ion{C}{4}   &   y91   &   \phm{x}   \nl
\phm{x}   &   \phm{x}   &   1.0472   &   \ion{Mg}{2}   &   0.4-1.7   &   \ion{Mg}{2}   &   y91   &   \phm{x}   \nl
\phm{x}   &   \phm{x}   &   1.1417   &   \ion{Mg}{2}   &   \phm{x}   &   \phm{x}   &   y91   &   \phm{x}   \nl
\phm{x}   &   \phm{x}   &   1.8332   &   \ion{C}{4}   &   \phm{x}   &   \phm{x}   &   btt90   &   Y   \nl
\phm{x}   &   \phm{x}   &   1.8753   &   \ion{C}{4}   &   \phm{x}   &   \phm{x}   &   btt90   &   Y   \nl
0831+101   &   1.76   &   1.7589   &   \ion{C}{4}   &   1.6-1.76   &   \ion{C}{4}   &   y91   &   Y   \nl
\phm{x}   &   \phm{x}   &   \phm{x}   &   \phm{x}   &   0.4-1.7   &   \ion{Mg}{2}   &   y91   &   \phm{x}   \nl
0835+580   &   1.534   &   1.4353   &   \ion{Mg}{2}   &   1.25-1.55   &   \ion{C}{4}   &   jhb91   &   \phm{x}   \nl
\phm{x}   &   \phm{x}   &   1.4383   &   \ion{Mg}{2}   &   0.22-1.55   &   \ion{Mg}{2}   &   btt90   &   \phm{x}   \nl
\phm{x}   &   \phm{x}   &   1.5322   &   \ion{C}{4}   &   \phm{x}   &   \phm{x}   &   jhb91   &   Y   \nl
\phm{x}   &   \phm{x}   &   1.5347   &   \ion{C}{4}   &   \phm{x}   &   \phm{x}   &   jhb91   &   Y   \nl
\phm{x}   &   \phm{x}   &   1.5431\tablenotemark{a}   &   \ion{Mg}{2},\ion{C}{4}   &   \phm{x}   &   \phm{x}   &   jhb91,abf97   &   Y   \nl
\tablebreak
0926+117   &   1.75   &   none   &   \phm{x}   &   1.70-1.75   &   \ion{C}{4}   &   btt90   &   \phm{x}   \nl
\phm{x}   &   \phm{x}   &   \phm{x}   &   \phm{x}   &   0.4-1.75   &   \ion{Mg}{2}   &   btt90   &   \phm{x}   \nl
0952+179   &   1.472   &   0.2378   &   \ion{Mg}{2}   &   1.0-1.5   &   \ion{C}{4}   &   y91   &   \phm{x}   \nl
\phm{x}   &   \phm{x}   &   \phm{x}   &   \phm{x}   &   0.11-1.5   &   \ion{Mg}{2}   &   \phm{x}   &   \phm{x}   \nl
1018+348   &   1.404   &   1.29   &   \ion{C}{4}   &   1.05-1.45   &   \ion{C}{4}   &   fun   &   \phm{x}   \nl
\phm{x}   &   \phm{x}   &   $\sim$1.403   &   \ion{C}{4}   &   0.14-0.78   &   \ion{Mg}{2}   &   fun   &   Y   \nl
\phm{x}   &   \phm{x}   &   $\sim$1.406   &   \ion{C}{4}   &   \phm{x}   &   \phm{x}   &   fun   &   Y   \nl
1126+101   &   1.516   &   1.4389   &   \ion{C}{4}   &   1.1-1.5   &   \ion{C}{4}   &   y91   &   \phm{x}   \nl
\phm{x}   &   \phm{x}   &   1.5098   &   \ion{C}{4}   &   0.14-0.52  &   \ion{Mg}{2}   &   y91   &   Y   \nl
\phm{x}   &   \phm{x}   &   1.5173   &   \ion{C}{4}   &   \phm{x}   &   \phm{x}   &   y91   &   Y   \nl
1218+339   &   1.52   &   0.7423   &   \ion{Mg}{2}   &   1.03-1.55   &   \ion{C}{4}   &   abe94   &   \phm{x}   \nl
\phm{x}   &   \phm{x}   &   1.5000   &   \ion{Mg}{2},\ion{C}{4}   &   0.12-1.55   &   \ion{Mg}{2}   &   abe94   &   Y   \nl
1221+113   &   1.755   &   1.6144   &   \ion{Mg}{2}   &   1.5-1.75   &   \ion{C}{4}   &   btt90   &   \phm{x}   \nl
\phm{x}   &   \phm{x}   &   \phm{x}   &   \phm{x}   &   0.4-1.75   &   \ion{Mg}{2}   &   \phm{x}   &   \phm{x}   \nl
1258+404   &   1.66   &   none   &   \phm{x}   &   1.55-1.65   &   \ion{C}{4}   &   btt90   &   \phm{x}   \nl
\phm{x}   &   \phm{x}   &   \phm{x}   &   \phm{x}   &   0.4-1.65   &   \ion{Mg}{2}   &   btt90   &   \phm{x}   \nl
1416+067   &   1.430   &   1.2734   &   \ion{C}{4}   &   1.2-1.6   &   \ion{C}{4}   &   y91   &   \phm{x}   \nl
\phm{x}   &   \phm{x}   &   1.3751   &   \ion{C}{4}   &   0.14-0.53   &   \ion{Mg}{2}   &   y91   &   \phm{x}   \nl
\phm{x}   &   \phm{x}   &   1.4348   &   \ion{C}{4}   &   \phm{x}   &   \phm{x}   &   y91   &   Y   \nl
\phm{x}   &   \phm{x}   &   1.4380   &   \ion{C}{4}   &   \phm{x}   &   \phm{x}   &   y91   &   Y   \nl
\phm{x}   &   \phm{x}   &   1.4408   &   \ion{C}{4}   &   \phm{x}   &   \phm{x}   &   y91   &   Y   \nl
1556+335   &   1.646   &   1.2321   &   \ion{C}{4}   &   1.1-1.65   &   \ion{C}{4}   &   y91   &   \phm{x}   \nl
\phm{x}   &   \phm{x}   &   1.6030   &   \ion{C}{4}   &   0.15-1.65   &   \ion{Mg}{2}   &   jhb91   &   Y   \nl
\phm{x}   &   \phm{x}   &   1.6106   &   \ion{C}{4}   &   \phm{x}   &   \phm{x}   &   jhb91   &   Y   \nl
\phm{x}   &   \phm{x}   &   1.6129   &   \ion{C}{4}   &   \phm{x}   &   \phm{x}   &   jhb91   &   Y   \nl
\phm{x}   &   \phm{x}   &   1.6395   &   \ion{C}{4}   &   \phm{x}   &   \phm{x}   &   jhb91   &   Y   \nl
\phm{x}   &   \phm{x}   &   1.6445   &   \ion{C}{4}   &   \phm{x}   &   \phm{x}   &   jhb91   &   Y   \nl
\phm{x}   &   \phm{x}   &   1.6505   &   \ion{Mg}{2},\ion{C}{4}   &   \phm{x}   &   \phm{x}   &   jhb91   &   Y   \nl
\phm{x}   &   \phm{x}   &   1.6519   &   \ion{Mg}{2},\ion{C}{4}   &   \phm{x}   &   \phm{x}   &   jhb91   &   Y   \nl
\phm{x}   &   \phm{x}   &   1.6537   &   \ion{Mg}{2},\ion{C}{4}   &   \phm{x}   &   \phm{x}   &   jhb91   &   Y   \nl
2044$-$168   &   1.939   &   1.3285   &   \ion{Mg}{2}   &   1.45-1.95   &   \ion{C}{4}   &   y91   &   \phm{x}   \nl
\phm{x}   &   \phm{x}   &   1.5586   &   \ion{C}{4}   &   0.35-0.75   &   \ion{Mg}{2}   &   y91   &   \phm{x}   \nl
\phm{x}   &   \phm{x}   &   1.7325   &   \ion{C}{4}   &   1.2-1.95   &   \ion{Mg}{2}   &   y91   &   \phm{x}   \nl
\phm{x}   &   \phm{x}   &   1.7341   &   \ion{C}{4}   &   \phm{x}   &   \phm{x}   &   y91   &   \phm{x}   \nl
\phm{x}   &   \phm{x}   &   1.7355   &   \ion{C}{4}   &   \phm{x}   &   \phm{x}   &   y91   &   \phm{x}   \nl
\phm{x}   &   \phm{x}   &   1.9199   &   \ion{C}{4}   &   \phm{x}   &   \phm{x}   &   y91   &   Y   \nl
\phm{x}   &   \phm{x}   &   1.9213   &   \ion{C}{4}   &   \phm{x}   &   \phm{x}   &   y91   &   Y   \nl
2149+212   &   1.5359   &   0.9114   &   \ion{Mg}{2}   &   1.5-1.55   &   \ion{C}{4}   &   y91   &   \phm{x}   \nl
\phm{x}   &   \phm{x}   &   1.0073   &   \ion{Mg}{2}   &   0.4-1.55   &   \ion{Mg}{2}   &   y91   &   \phm{x}   \nl
2345+061   &   1.54   &   none   &   \phm{x}   &   1.50-1.55   &   \ion{C}{4}   &   btt90   &   \phm{x}   \nl
\phm{x}   &   \phm{x}   &   \phm{x}   &   \phm{x}   &   0.4-1.55   &   \ion{Mg}{2}   &   \phm{x}   &   \phm{x}   \nl
\enddata
\tablecomments{Only \ion{C}{4} or \ion{Mg}{2} absorbers are listed, one
absorber per line.  Also listed is the redshift search interval, the redshift
range which has been searched for \ion{C}{4} or \ion{Mg}{2} absorption, as
indicated.  ``Ass. Abs.?'' refers to whether or not the particular absorber is
an associated absorber.
}
\tablenotetext{a}{Aldcroft, Bechtold \& Foltz (1997) report variability in this
system (measured at $z$=1.5425, not 1.5431, in their spectra) over 3.9 years
rest frame.  This means this absorption system is very probably intrinsic to
the quasar.}
\tablerefs{abe94: Aldcroft, Bechtold \& Elvis (1994);
abf97: Aldcroft, Bechtold \& Elvis (1994);
bah93: Bahcall {\it et~al.} (1993);
btt90: Barthel, Tytler \& Thomson (1990);
bt96: Burles \& Tytler (1996);
fcw88: Foltz, Chaffee \& Wolfe (1988);
fun: Foltz {\it et al.}, unpublished;
jhb91: Junkkarinen, Hewitt \& Burbidge (1991), see also Junkkarinen, Hewitt \& Burbidge (1992);
ssh97: Sowinski, Schmidt \& Hines (1997);
ss92: Steidel \& Sargent (1992);
wil95: Wills {\it et~al.} (1995);
y91: York {\it et~al.} (1991).
}
\end{deluxetable}

\begin{deluxetable}{lllcccccc}    
\tablecaption{Extinction, Magnitude Limits, and Seeing in Observed Fields\label{tab_zg1datainfo}}
\tablewidth{432.00000pt}
\tablenum{4}
\tablehead{
\colhead{} & \colhead{Gal.} & \colhead{5$\sigma$} &
\multicolumn{2}{c}{$K_s$ band} & \multicolumn{2}{c}{$r$ band} &
\colhead{Other filters} & \colhead{Area,} \\[.2ex]
\colhead{Field} & \colhead{Ext.} & \colhead{$K_s$ lim} &
\colhead{3$\sigma$ lim} & \colhead{seeing} & \colhead{3$\sigma$ lim} & \colhead{seeing} &
\colhead{3$\sigma$ lim, seeing} & \colhead{arcmin$^2$}}
\startdata
Q~0003$-$003\tablenotemark{d} & 0.0132 & 20.0\tablenotemark{a}  & 20.823 & 1.51 & 25.285 & 1.67 & \nodata & 6.894 \\
Q~0017+154\tablenotemark{d} & 0.0253 & 20.962 & 21.512 & 1.09 & 25.890 & 1.57 & \nodata & 7.684 \\
Q~0033+098   & 0.0472 & 21.006 & 21.556 & 1.51 & 25.363 & 1.39 & \nodata & 5.480 \\
Q~0149+218   & 0.0612 & 19.724\tablenotemark{a} & 20.307 & 1.21 & 25.760 & 1.66 & \nodata & 7.261 \\
Q~0232$-$042 & 0.0053 & 19.5   & 21.470 & 1.25 & 25.540 & 1.72 & \nodata & 7.598 \\
Q~0256$-$005 & 0.0712 & 20.906 & 21.456 & 1.21 & 24.450 & 1.39 & \nodata & 5.680 \\
Q~0352+123   & 0.1892 & 19.0   & 20.344 & 1.45 & 24.985 & 1.51 & \nodata & 6.413 \\
Q~0736$-$063 & 0.27\tablenotemark{b} & \nodata & 21.205 & 1.12 & 24.177 & 1.39 & \nodata & 5.493 \\
Q~0808+289   & 0.0322 & 20.906\tablenotemark{c} & 21.456 & 1.51 & 25.330 & 1.55 & \nodata & 6.835 \\
Q~0831+101   & 0.0333 & 21.513\tablenotemark{c} & 22.063 & 1.24 & 25.501 & 1.18 & \nodata & 6.815 \\
Q~0835+580\tablenotemark{d} & 0.0522 & 21.241 & 21.791 & 1.21 & 25.838 & 1.28 & $H$:20.896,1.27 & 8.527 \\
            &        &        &        &      &        &      & $J$:23.689,1.37 &       \\
            &        &        &        &      &        &      & $z$:24.990,1.64 &       \\
            &        &        &        &      &        &      & $i$:24.928,1.18 &       \\
Q~0926+117   & 0.0153 & 20.689 & 21.239 & 1.45 & 25.447 & 1.12 & \nodata & 5.568 \\
Q~0952+179   & 0.0203 & 21.230 & 21.780 & 1.60 & 25.622 & 1.21 & $J$:23.807,1.39 & 8.613 \\
Q~1018+348   & 0.0000 & 20.359 & 20.909 & 1.48 & 25.705 & 1.27 & \nodata & 7.605 \\
Q~1126+101   & 0.0263 & 20.788 & 21.338 & 1.15 & 25.877 & 1.36 & $H$:20.939,1.12 & 8.585 \\
            &        &        &        &      &        &      & $J$:23.226,1.46 &       \\
            &        &        &        &      &        &      & $z$:25.009,1.87 &       \\
Q~1218+339\tablenotemark{d} & 0.0002 & 20.677 & 21.227 & 1.42 & 25.529 & 1.21 & \nodata & 7.936 \\
Q~1221+113   & 0.0072 & 20.864 & 21.414 & 1.22 & 25.599 & 1.33 & \nodata & 7.157 \\
Q~1258+404\tablenotemark{d} & 0.0000 & 21.613\tablenotemark{c} & 22.163 & 1.45 & 25.818 & 1.15 & $J$:23.479,1.66 & 9.199 \\
            &        &        &        &      &        &      & $I$:23.923,1.54 &       \\
Q~1328+254\tablenotemark{d} & 0.0062 & 19.0   & 20.099 & 2.20 & 25.844 & 1.36 & \nodata & 6.695 \\
Q~1416+067\tablenotemark{d} & 0.0072 & 20.691 & 21.241 & 1.30 & 25.812 & 1.45 & \nodata & 7.071 \\
Q~1430-0046  & 0.0233 & 19.229\tablenotemark{a} & 19.785 & 1.66 & 25.224 & 1.51 & \nodata & 7.028 \\
Q~1437+624   & 0.0012 & 18.5\tablenotemark{c}  & 20.245 & 1.30 & 25.318 & 1.33 & \nodata & 7.673 \\
Q~1508$-$055 & 0.0542 & \nodata & \nodata & 1.91 & \nodata & \nodata & \nodata & \nodata \\
Q~1556+335   & 0.0183 & 20.169\tablenotemark{a} & 20.719 & 1.25 & 25.593 & 1.18 & \nodata & 7.174 \\
Q~1606+106   & 0.0432 & 20.0\tablenotemark{c}  & 20.753 & 1.43 & 25.420 & 2.23 & \nodata & 5.543 \\
Q~1718+481   & 0.0213 & 20.0   & 20.530 & 1.45 & 23.925 & 1.75 & \nodata & 7.446 \\
Q~1739+522   & 0.0333 & 19.0\tablenotemark{c}  & 20.704 & 1.27 & 25.776 & 1.24 & \nodata & 6.258 \\
Q~2044$-$168 & 0.0492 & 20.148\tablenotemark{a} & 20.698 & 1.21 & 25.408 & 1.69 & \nodata & 5.502 \\
Q~2144+092   & 0.0542 & 18.0\tablenotemark{c}  & 21.138 & 1.78 & 25.098 & 1.69 & \nodata & 7.593 \\
Q~2149+212   & 0.0822 & 20.639 & 21.189 & 1.54 & 25.558 & 1.27 & \nodata & 6.690 \\
Q~2230+114   & 0.0422 & 20.168\tablenotemark{a} & 20.749 & 1.63 & 25.327 & 1.66 & \nodata & 7.316 \\
Q~2325+293   & 0.0702 & 20.164 & 20.742 & 1.39 & 25.424 & 1.28 & \nodata & 6.012 \\
Q~2345+061   & 0.0653 & 21.169 & 21.719 & 1.58 & 25.111 & 1.42 & $J$:23.072,2.68 & 9.347 \\
CtrlFld1    & 0.0273 & 20.525 & 21.075 & 1.12 & 24.738 & 1.05 & \nodata & 9.435 \\
CtrlFld3    & 0.0192 & 20.726 & 21.276 & 1.60 & 24.743 & 1.05 & \nodata & 9.343 \\
\enddata
\tablenotetext{a}{3$\sigma_{sky}$ limit above sky used instead of 2.5$\sigma_{sky}$ (see \S\ref{detect}).}
\tablenotetext{b}{Assumed value.}
\tablenotetext{c}{Minimum detection area larger than 1.9 arcsec$^2$ used (see \S\ref{detect}).}
\tablenotetext{d}{$HST$ archival WFPC2 data of useful depth available (see \S\ref{classify}).}
\tablecomments{Galactic extinction ``Gal. Ext.'' is $E$($B$$-$$V$).
5$\sigma$ $K_s$ limit magnitudes with only three significant digits are fields 
where, for various regions, detections are not reliable down to the nominal 
5$\sigma$ limits, but only to the values listed (see \S\ref{detect}).
Seeing measurements are in arcseconds.
Area given is the overlapping area between $r$ and $K_s$ images,
as well as $J$ for all fields with $J$ data except Q~2345+061.}
\end{deluxetable}

\begin{deluxetable}{ccccccrc}	
\tablecaption{$K_{UKIRT}$ Galaxy Counts:  Our Control Fields\label{tab_knm_cf}}
\tablewidth{432.00000pt}
\tablenum{5}
\tablehead{
\colhead{} & \colhead{} &
\multicolumn{2}{c}{Poisson Uncertainties} &
\multicolumn{2}{c}{RMS Uncertainties} &
\colhead{Area} & \colhead{Number} \\[.2ex]
\colhead{$K$} & \colhead{log~$N$} &
\colhead{Lower} & \colhead{Upper} &
\colhead{Lower} & \colhead{Upper} &
\colhead{(arcmin$^2$)} & \colhead{of Fields}}
\startdata
15.25 & 2.585 & 0.448 & 0.367 & 2.585 & 0.384 & 18.778 & 2.0 \\
15.75 & 2.888 & 0.296 & 0.354 & 0.003 & 0.003 & 18.778 & 2.0 \\
16.25 & 3.190 & 0.210 & 0.248 & 0.544 & 0.234 & 18.778 & 2.0 \\
16.75 & 3.067 & 0.339 & 0.296 & 3.067 & 0.382 & 18.778 & 2.0 \\
17.25 & 3.704 & 0.119 & 0.125 & 0.610 & 0.244 & 18.778 & 2.0 \\
17.75 & 3.494 & 0.150 & 0.162 & 0.003 & 0.003 & 18.778 & 2.0 \\
18.25 & 3.592 & 0.150 & 0.162 & 0.003 & 0.003 & 18.778 & 2.0 \\
18.75 & 3.854 & 0.105 & 0.111 & 0.333 & 0.186 & 18.778 & 2.0 \\
19.25 & 4.227 & 0.069 & 0.072 & 0.169 & 0.121 & 18.778 & 2.0 \\
19.75 & 4.245 & 0.069 & 0.073 & 0.019 & 0.018 & 18.559 & 2.0 \\
20.25 & 4.348 & 0.067 & 0.070 & 0.065 & 0.057 & 16.718 & 2.0 \\
20.75 & 4.614 & 0.063 & 0.066 & 0.302 & 0.176 & 12.219 & 2.0 \\
\enddata
\tablecomments{Units of $N$ are number mag$^{-1}$ deg$^{-2}$.
Number of fields refers to the two separate control field pointings
used to calculate the RMS errors.}
\end{deluxetable}

\begin{deluxetable}{ccccccrc}	
\tablecaption{$K_{UKIRT}$ Galaxy Counts:  Literature Average\label{tab_knm_puball}}
\tablewidth{432.00000pt}
\tablenum{6}
\tablehead{
\colhead{} & \colhead{} &
\multicolumn{2}{c}{Poisson Uncertainties} &
\multicolumn{2}{c}{RMS Uncertainties} &
\colhead{Area} & \colhead{Number of} \\[.2ex]
\colhead{$K$} & \colhead{log~$N$} &
\colhead{Lower} & \colhead{Upper} &
\colhead{Lower} & \colhead{Upper} &
\colhead{(arcmin$^2$)} & \colhead{Surveys}}
\startdata
12.75 & 1.379 & 0.451 & 0.366 & 0.108 & 0.086 & 600.808 & 3.0 \\
13.25 & 0.816 & 0.739 & 0.558 & 0.816 & 0.446 & 781.608 & 4.0 \\
13.75 & 1.519 & 0.162 & 0.214 & 0.516 & 0.229 & 1668.088 & 7.0 \\
14.25 & 1.708 & 0.144 & 0.168 & 0.304 & 0.177 & 1668.088 & 7.0 \\
14.75 & 2.136 & 0.103 & 0.101 & 0.159 & 0.116 & 1668.088 & 7.0 \\
15.25 & 2.462 & 0.071 & 0.061 & 0.108 & 0.086 & 1668.088 & 7.0 \\
15.75 & 2.772 & 0.039 & 0.038 & 0.118 & 0.093 & 1684.598 & 8.0 \\
16.25 & 3.001 & 0.051 & 0.029 & 0.102 & 0.082 & 1684.598 & 8.0 \\
16.75 & 3.279 & 0.058 & 0.025 & 0.075 & 0.064 & 1196.398 & 9.0 \\
17.25 & 3.534 & 0.027 & 0.026 & 0.183 & 0.128 & 565.878 & 8.0 \\
17.75 & 3.699 & 0.031 & 0.022 & 0.147 & 0.109 & 565.878 & 8.0 \\
18.25 & 3.889 & 0.019 & 0.019 & 0.179 & 0.126 & 511.638 & 9.0 \\
18.75 & 4.077 & 0.027 & 0.027 & 0.125 & 0.097 & 294.928 & 9.0 \\
19.25 & 4.231 & 0.041 & 0.037 & 0.104 & 0.084 & 278.998 & 8.0 \\
19.75 & 4.341 & 0.051 & 0.056 & 0.104 & 0.084 & 59.499 & 6.0 \\
20.25 & 4.442 & 0.046 & 0.046 & 0.156 & 0.115 & 33.898 & 7.0 \\
20.75 & 4.639 & 0.040 & 0.044 & 0.164 & 0.119 & 27.609 & 7.0 \\
21.25 & 4.782 & 0.049 & 0.054 & 0.172 & 0.123 & 13.590 & 5.0 \\
21.75 & 4.913 & 0.048 & 0.057 & 0.127 & 0.098 & 7.380 & 3.0 \\
22.25 & 5.032 & 0.041 & 0.052 & 0.147 & 0.110 & 5.370 & 3.0 \\
22.75 & 5.172 & 0.052 & 0.057 & 0.131 & 0.101 & 4.210 & 2.0 \\
23.25 & 5.367 & 0.057 & 0.060 & 0.096 & 0.079 & 4.210 & 2.0 \\
23.75 & 5.597 & 0.085 & 0.085 & 0.095 & 0.078 & 2.880 & 1.0 \\
\enddata
\tablecomments{Units of $N$ are number mag$^{-1}$ deg$^{-2}$.
Number of surveys refers to the number of separate surveys used to calculate
the RMS errors.}
\end{deluxetable}

\end{document}